\documentclass[12pt,preprint]{aastex}

\usepackage{natbib}
\usepackage{epsfig,graphicx,color}
\usepackage{longtable}

\definecolor{orange}{cmyk}{0,0.4,0.8,0.2}
\definecolor{darkorange}{rgb}{.71,0.21,0.01}
\definecolor{darkgreen}{rgb}{.12,.54,.11}

\usepackage{hyperref}
\hypersetup{pdftex,  
  breaklinks=true,  
  colorlinks=true,
  urlcolor=blue,
  linkcolor=darkorange,
  citecolor=darkgreen,
  }

\newcommand{\x}{{\bf x}}

\newcommand{\z}{{\bf z}}

\newcommand{\Prob}{\mathbf{P}}
\newcommand{\PrfL}{\widehat{P}_{\rm{RF,}\mathcal{L}}(y|\x)}
\newcommand{\PrfLx}{\widehat{P}_{\rm{RF,}\mathcal{L}\cup\x'}(y|\x)}
\newcommand{\TreeL}{\theta_{b, \mathcal{L}}(y|\x)}
\newcommand{\TreeLx}{\theta_{b, \mathcal{L}\cup\x'}(y|\x)}
 \def\hipp {{\it Hipparcos~}}

\begin{document}

\shorttitle{Active Learning for Variable Star Classification}
\shortauthors{J. W. Richards, et al.}
\title{Active Learning to Overcome Sample Selection Bias: Application to Photometric Variable Star Classification}
\author{
Joseph W. Richards\altaffilmark{1,2,*},
Dan L. Starr\altaffilmark{1},
Henrik Brink\altaffilmark{3},
Adam A. Miller\altaffilmark{1},
Joshua S. Bloom\altaffilmark{1},
Nathaniel R. Butler\altaffilmark{1},
J. Berian James\altaffilmark{1,3},
James P. Long\altaffilmark{2}
and John Rice\altaffilmark{2}
}

\altaffiltext{1}{Astronomy Department, University of California, Berkeley, CA, 94720-7450, USA}
\altaffiltext{2}{Statistics Department, University of California, Berkeley, CA, 94720-7450, USA}
\altaffiltext{3}{Dark Cosmology Centre, Juliane Maries Vej 30, 2100 Copenhagen \O, Denmark }
\altaffiltext{*}{E-mail: {\tt jwrichar@stat.berkeley.edu}}

\slugcomment{Submitted}

\begin{abstract}
Despite the great promise of machine-learning algorithms to classify and predict astrophysical parameters for the vast numbers of astrophysical sources and transients observed in large-scale surveys, the peculiarities of the training data often manifest as strongly biased predictions on the data of interest. Typically, training sets are derived from historical surveys of brighter, more nearby objects than those from more extensive, deeper surveys (\emph{testing data}). This \emph{sample selection bias} can cause catastrophic errors in predictions on the testing data because a) standard assumptions for machine-learned model selection procedures break down and b) dense regions of testing space might be completely devoid of training data.  We explore possible remedies to sample selection bias, including importance weighting (IW), co-training (CT), and active learning (AL).  We argue that AL---where the data whose inclusion in the training set would most improve predictions on the testing set are queried for manual follow-up---is an effective approach and is appropriate for many astronomical applications.  For a variable star classification problem on a well-studied set of stars from \hipp and OGLE, AL is the optimal method in terms of error rate on the testing data, beating the off-the-shelf classifier by  3.4\% and the other proposed methods by at least 3.0\%.  To aid with manual labeling of variable stars, we developed a web interface which allows for easy light curve visualization and querying of external databases.  Finally, we apply active learning to classify variable stars in the ASAS survey, finding dramatic improvement in our agreement with the ACVS catalog, from 65.5\% to 79.5\%, and a significant increase in the classifier's average confidence for the testing set, from 14.6\% to 42.9\%, after a few AL iterations.
\end{abstract}


\section{Introduction}
\label{sec:intro}

Automated classification and parameter estimation procedures are crucial for the analysis of upcoming astronomical surveys.  Planned missions such as Gaia (\citealt{2001A&A...369..339P}) and the Large Synoptic Survey Telescope (LSST; \citealt{2009arXiv0912.0201L}) will collect data for more than a billion objects, making it impossible for researchers to manually study significant subsets  of the data.  At the same time, these upcoming missions will probe never-before-seen regions of astrophysical parameter space and will do so with larger telescopes and more precise detectors.  This makes  the training of automated learners for these new surveys  a difficult, non-trivial task.

Supervised machine learning methods (see \citet{br11} for review) have shown great promise for the automatic estimation of astrophysical quantities of interest---\emph{response} variables in the statistics parlance---from sets of \emph{features} extracted from the observed data.  These studies include areas as diverse as photometric redshift estimation (\citealt{2004PASP..116..345C}, \citealt{2005PASP..117...79W}, \citealt{2007ApJ...663..752D}, \citealt{2010ApJ...712..511C}), stellar parameter estimation and classification (\citealt{2007A&A...470..761T}, \citealt{2010A&A...522A..88S}), galaxy morphology classification (\citealt{2004MNRAS.348.1038B}, \citealt{2008A&A...478..971H}), galaxy-star separation (\citealt{2008MNRAS.386.1417G}, \citealt{2009AJ....137.3884R}), supernova typing (\citealt{2011MNRAS.tmp..545N}, \citealt{2011arXiv1103.6034R}) and variable star classification (\citealt{2007debo}, \citealt{2011arXiv1101.2406D}, \citealt{2011rich}), among others. 

These studies typically assume that the distribution of training data is representative of the set of data to be analyzed (the so-called \emph{testing} data).  In reality, in astronomy the distributions of  training and testing data are usually substantially different.  This \emph{sample selection bias} can cause significant problems for an automated supervised method and must be addressed to ensure satisfactory performance for the testing data.  For instance, standard cross-validation techniques assume that the training and testing distributions are exactly the same; when this is not the case, sub-optimal model selection can occur.  

In this paper, we show the debilitating effects of sample selection bias on the problem of automated classification of variable stars from their observed light curves.  Using a set of highly studied, well-classified variable star light curves from the \hipp (\citealt{1997perr}) Space Astrometry Mission and the Optical Gravitational Lensing Experiment (OGLE, \citealt{1999udal}) missions, we train a classifier to automatically predict the class of each variable star in the All Sky Automated Survey (ASAS, \citealt{1997AcA....47..467P}, \citealt{2001ASPC..246...53P}).  We demonstrate that this classifier results in a high error rate, a substantial number of anomalies, and low average classifier confidence.  These debilitating effects are also seen in existing catalogs such as the ACVS (\citealt{2000AcA....50..177P}, \citealt{acvs}), whose use of training data from OGLE plus from an early ASAS release yields a supervised classifier that is only confident on 24\% of all sources.   Upcoming surveys, whose automated prediction algorithms will be trained on data from older surveys or idealized models, will suffer from these same maladies if sample selection bias is not treated properly.

To overcome sample selection bias, we propose a few methods, including importance weighting, co-training, and active learning.  On both the ASAS variable star classification problem and a simulated variable star data set, we find that active learning (AL) performs the best.  AL is an iterative procedure, whereby on each iteration the testing data whose inclusion in the training set would most improve predictions over the entire testing set are queried for manual follow-up and added to the training set.   AL is a semi-supervised method that leverages the known features of the testing data to make the best decision about which of these objects is most useful to the supervised learner.  We argue that active learning is appropriate in many areas of astrophysics, where follow-up information can often be attained through spectroscopic observations, manual study, or citizen science projects (e.g., \citealt{2008MNRAS.389.1179L}).  Furthermore, AL is a principled method for selecting objects for expensive follow-up in circumstances where it is infeasible to perform an in-depth analysis on every object.  In particular, projects such as Galaxy Zoo stand to benefit from the active learning approach for candidate object selection, especially when data sizes become prohibitively large to manually analyze each source.

The structure of the paper is  as follows.  In \S\ref{sec:sampBias} we describe in detail the problem of sample selection bias, showing how it can arise in various astronomical settings and detailing its adverse effects in a variable star classification problem.  In \S\ref{sec:methods} we propose a few methods that can be used to mitigate the effects of sample selection bias.  We describe active learning in detail, focusing on its implementation with Random Forest classification.  Next, we test those methods in \S\ref{sec:deb}, showing that AL attains the best results in a simulated variable star classification experiment.  In \S\ref{sec:allstars} we describe our online active learning variable star classification tool, {\tt ALLSTARS}, which was developed to aid the manual study of objects in various photometric surveys. We present the result of applying active learning to classify ASAS variable stars in \S\ref{sec:results}, showing drastic improvement over the off-the-shelf classifier.  Finally, we end with some concluding remarks in \S\ref{sec:conclusions}.

\section{Sample Selection Bias in Astronomical Surveys}
\label{sec:sampBias}

A fundamental assumption for supervised machine learning methods is that the training and testing sets\footnote{Throughout the paper, we  call training data those objects with known response variable that are used to train the supervised model, and we call testing data the objects of interest whose unknown response is to be predicted by the model.} are drawn independently from the same underlying distribution.  However, in astrophysics this is rarely the case. Populations of well-understood, well-studied training objects are inherently biased toward intrinsically brighter and nearby sources and available data are typically from older, lower signal-to-noise detectors.   

Indeed, in studies of variable stars, samples of more luminous, well-understood stars are often employed to train supervised algorithms to classify fainter stars observed by newer, deeper surveys.  Examples of this abound in the literature.  For instance, \citet{2009A&A...506..519D} use a training set from OGLE, a ground-based survey from Las Campanas Observatory covering fields in the Magellanic Clouds and Galactic bulge, to classify higher-quality CoRoT (COnvection ROtation and planetary Transits, \citealt{2009A&A...506..411A}) satellite data.   \citet{2011arXiv1101.2406D} train a classification model using a subset of the \hipp periodic star catalog containing the most reliable labels from the literature and most confident period estimates. This systematic difference between the training and testing sets can cause supervised methods to perform poorly, especially for the types of object under-sampled by the training set.  

\begin{figure}
\begin{center}
\includegraphics[angle=0,width=6.5in]{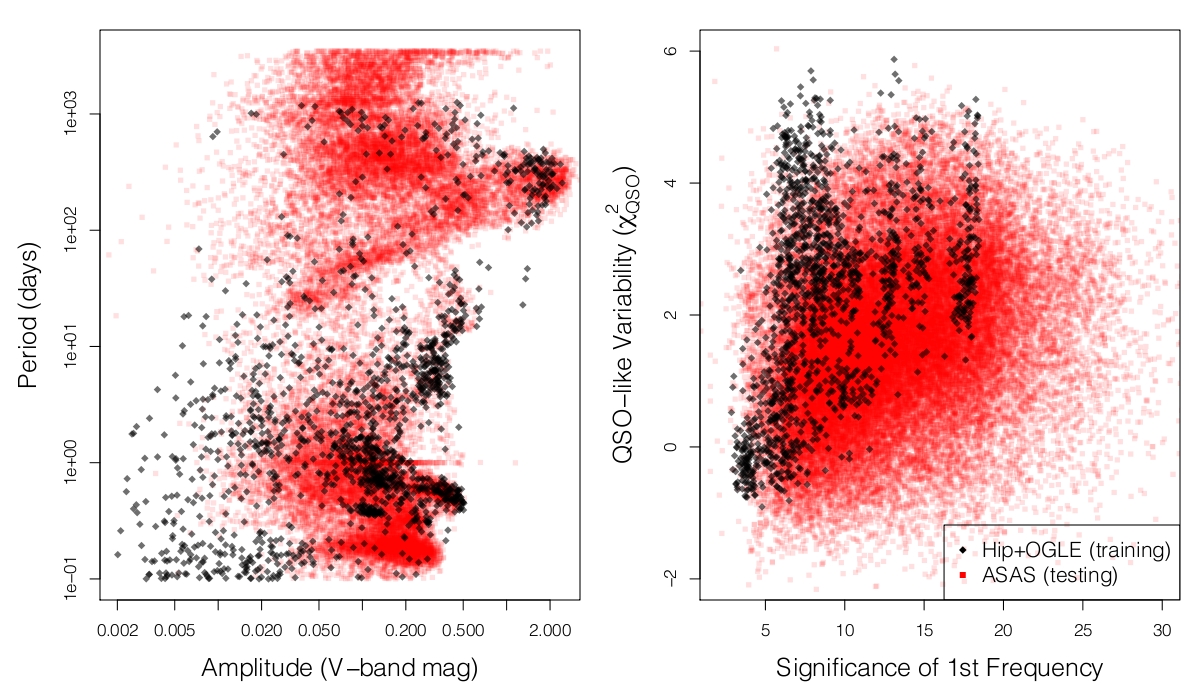}
\end{center}
\caption{ Sample selection bias for ASAS variable star (red $\Box$) classification using a training set of well-understood data from \hipp and OGLE (black $\diamond$).  Left: Large distributional mismatch exists in the period-amplitude plane.  Only those ASAS data whose statistical significance of the frequency estimate is larger than the median are plotted.  ASAS testing data have high density in short-period, high-amplitude and long-period, moderate-amplitude regions, where there are little training data.  Right: Testing data tend to have smaller values of the QSO-like variability metric---which measures how well the observed light curve fits a damped random walk QSO model (see \citealt{2011butl})---and larger values of the statistical significance of the first frequency (compared to a null, white-noise model; see \citealt{2011rich}). \label{fig:traintestoffset} }
\end{figure}

In \citet{2009A&A...506..519D}, the authors recognize that a training set ``should be constructed from data measured with the same instrument as the data to be classified" and claim that some misclassifications occur in their analysis due to systematic differences between the two surveys.  Because the aims and specifications of each survey are different, their observed sources usually occupy different regions of feature space.  See, for example, Fig. \ref{fig:traintestoffset}, where there is an obvious absence of the combined \hipp and OGLE training data in the high-frequency, high-amplitude regime where the density of the testing set of ASAS variables is high.  Even if two surveys have similar specifications (e.g., cadence, filter, depth), they may be looking in different parts of the sky or with different sensitivities and thus will observe different demographics of the same sources, causing a systematic differences in the survey priors.

In other areas of astrophysics and cosmology it is common practice to construct supervised models using spectroscopic samples and apply those models to predict parameters of interest for objects that fall entirely outside the support of the distribution of the spectroscopic data.  For example, photometric redshift estimation methods typically train a regression model using a set of spectroscopically confirmed objects, whereby those models are extended to populations of galaxies that are fainter and (often) at higher redshift (papers that have studied this problem include \citealt{2010MNRAS.405..987B} and \citealt{2011AAS...21715004S}).  Several authors have proposed novel methods to mitigate the effects of non-representative photo-z training sets using  physical association of galaxies (\citealt{2010ApJ...721..456M,2010ApJ...725..794Q}) or calibration through cross-correlation (\citealt{2010ApJ...724.1305S}).  Another field where these issues occur is supernova typing, where classifiers are typically trained on spectroscopically confirmed templates and then applied to classify fainter testing data (\citealt{2010kess,2011MNRAS.tmp..545N}).  Recently, \citet{2011arXiv1103.6034R} studied the impact of the accuracy of a supervised supernova classification method on the particular spectroscopic strategy employed to obtain training sets, finding that deeper samples with fewer objects are preferred to surveys with shallower limits.

The situation we describe, where the training and testing samples are generated from different distributions, is referred to in the statistics and machine learning literature as \emph{covariate shift} (\citealt{shimo2000}) or \emph{sample selection bias} (\citealt{heckman1979}).  This systematic difference can cause catastrophic prediction errors when the trained model is applied to new data.  These problems arise for two reasons.  First, under sample selection bias, standard generalization error estimation procedures, such as cross-validation, are biased, resulting in poor model selection.  Off-the-shelf supervised methods are designed to choose the model that minimizes some error criterion integrated with respect to the training distribution; when the testing distribution is substantially different, this model is likely to be suboptimal for prediction on the testing data.  In (\S \ref{ss:impwt}) we describe a principled weighting scheme to alleviate this complication.  Second,  significant regions of parameter space may be ignored by the training data---such as in the variable star classification problem shown in Fig. \ref{fig:traintestoffset}---causing catastrophically bad extrapolation of the model onto those regions.  In this case, any classifier trained only on the training data will produce poor class predictions for the ignored regions of parameter space: no weighting scheme on the training data can enforce good classifier performance in these regions.  This suggests that the testing data need to be used, in a semi-supervised manner,  to augment the training set.  In this paper, we explore two different approaches to this problem: \emph{co-training} (\S\ref{ss:cotrain} and \emph{self-training}), where testing instances with most certain class prediction are iteratively added to the training set, and \emph{active learning} (\S\ref{ss:actlearn}), where testing instances whose labels, if known, would be of maximal benefit to the supervised method, are manually studied to ascertain the value of their response (e.g. class label, redshift, etc.), and subsequently included in the training set.

\subsection{Example: Source Classification for ASAS}
\label{ss:acvs}

In this section, we demonstrate the effects of sample selection bias in classifying variable stars from the All Sky Automated Survey (ASAS). Particularly, we use an automated machine learning classifier to classify sources in the ASAS Catalogue of Variable Stars (ACVS, \citealt{2002AcA....52..397P}).  ACVS verson 1.1\footnote{The ACVS catalog can be downloaded at \url{http://www.astrouw.edu.pl/asas/data/ACVS.1.1.gz}.} consists of $V$-band light curves for 50,124 stars that have passed tests of variability as described in \citet{2000AcA....50..177P}.  As a training set for this classification problem, we use only the confidently labeled \hipp and OGLE sources used in \citet{2007debo} and \citet{2011rich}.  This data set consists of 1542 variable stars from 25 different science classes.  The period-amplitude relationship of the instances in the training set of \hipp and OGLE data, and in the ACVS catalog are plotted in Figure \ref{fig:traintestoffset}, where sample selection bias is obvious.

\begin{figure*}
\begin{center}
\includegraphics[angle=0,width=6.5in]{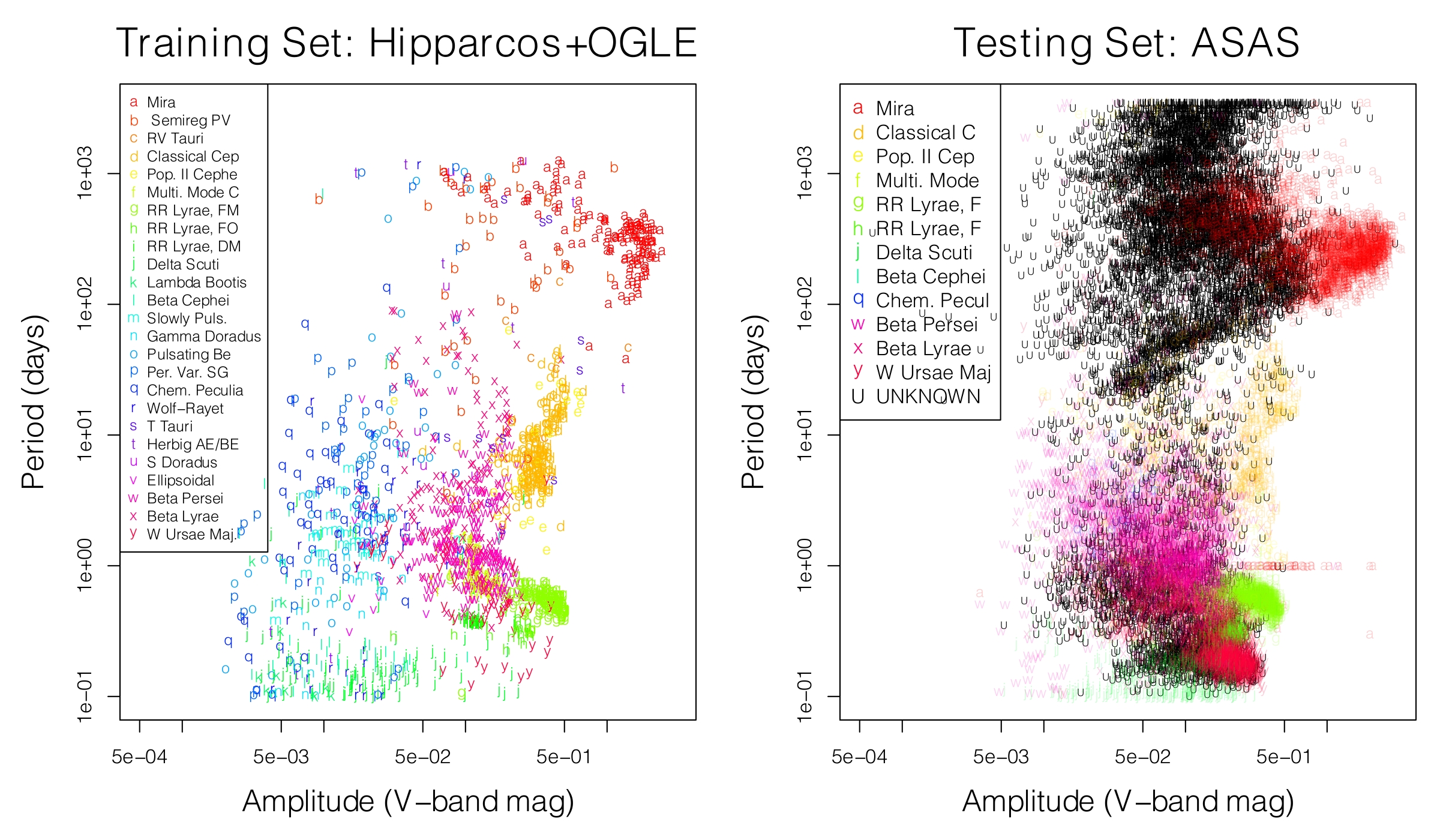}
\end{center}
\caption{ Left: Period-amplitude relationship for the 1542 training set sources from the \hipp and OGLE surveys.  Symbols and colors denote the true science class of each object.  Right:  Same for an arbitrary sample of size 10,000 from the 50,124 ASAS testing objects, where symbols and colors denote the ACVS labels.  Black `U' denotes that the source is either labeled MISC, doubly-labeled, or has low confidence label by ACVS.  Our goal is to use the training data set to predict the class label (and posterior class probabilities) for each ASAS object.  Complicating this task is the significant distributional difference between the training and testing sets.   \label{fig:traintestclass}}
\end{figure*}

As a part of ACVS, predicted classes are provided for a fraction of the stars.  As described in \citet{2002AcA....52..397P}, ACVS obtains their classifications using a neural net type algorithm trained on set of  visually labeled ASAS sources, confirmed OGLE cepheids \citet{1999AcA....49..223U,1999AcA....49..437U}, and OGLE Bulge variable stars \citet{2002AcA....52..129W}.   A filter is used to divide strictly periodic from less regular periodic sources.  A neural net is trained on the period, amplitude, Fourier coefficients (first 4 harmonics), $J-H$ and $H-K$ colors and IR fluxes to predict the  classes of the strictly periodic sources.  Several ACVS objects either have multiple labels or are annotated as having low confidence classifications.  For less regular periodic sources, location in the $J-H$ vs. $H-K$ plane is tested; if the object falls within an area of late-type irregular or semi-regular stars, it is assigned the label MISC, else it is inspected by eye.  We find that 38,117 ACVS stars, representing 76\% of the catalog,  are either labeled as MISC, assigned multiple labels, or have low class confidence.  The remaining 24\% of stars have confident ACVS labels, and provide a set of classifications to compare our algorithms against.  In Figure \ref{fig:traintestclass} we plot in color, in period-amplitude space, the classes of the training data and the ACVS classes of the ASAS data\footnote{Note that not all sources are actually periodic, meaning that some period estimates are nonsensical.  However, we also use the statistical significance of the frequency estimate as an input feature into our classifier; thus the classifier learns to  trust the only periodic features of those sources with high frequency significance, and to rely on only the non-periodic features of the low-significance data.}.

As our base model, we use a Random Forest classifier (\citealt{2001brei}).  Random Forest has recently been shown by \citet{2011rich} and \citet{2011arXiv1101.2406D} to attain accurate results in automated classification of variable stars.  In this paper, we represent each variable star in our data set by the 59 light-curve features used by \citet{2011rich}, as well as 5 additional light-curve features from \citet{2011arXiv1101.2406D}. The Random Forest classifier is a supervised, non-parametric  method that attempts to predict the science class of each star from its high-dimensional feature vector.  It  operates by constructing an ensemble of classification decision trees, and subsequently averaging the results. The key to the good performance of Random Forest is that its component trees are de-correlated by sub-selecting a small random number of features as splitting candidates in each non-terminal node of the tree. As a result, the average of the de-correlated trees attains highly decreased variance over each single tree, with no substantial increase in bias.\footnote{For more details about the Random Forest variable star classifier used, see \citet{2011rich}.}. 

\begin{figure*}
\begin{center}
\includegraphics[angle=0,width=6.5in]{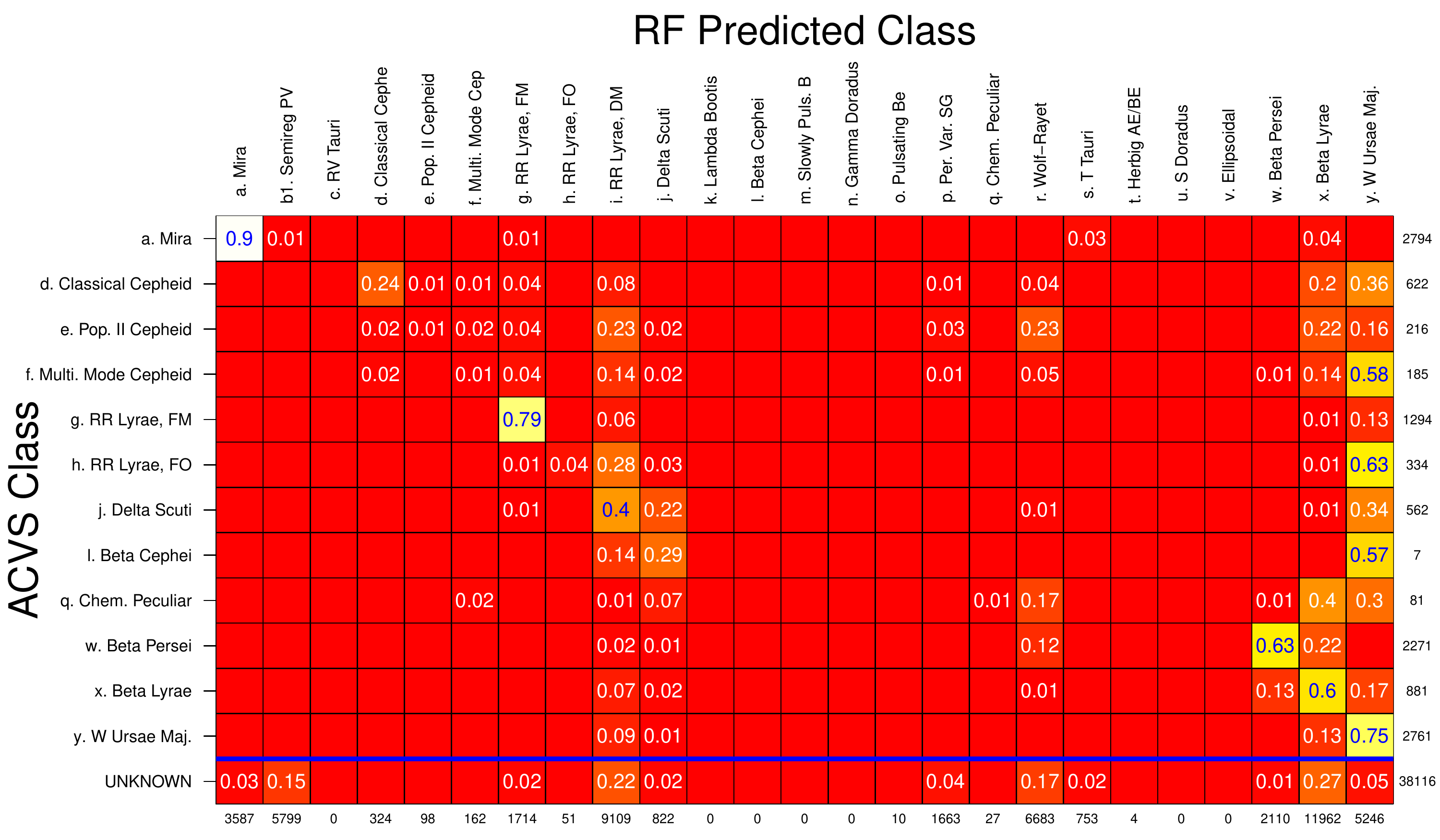}
\end{center}
\caption{Off-the-shelf Random Forest classifications of the ASAS data set, using a training set of the 1542 \hipp \& OGLE sources, compared to the ACVS classifications.  Rows are normalized to sum to 100\%, marginal counts are listed to the right and bottom of the table.  The RF classifier finds a 65.5\% correspondence with the ACVS labels, for the 12,007 objects with ACVS label, with many major discrepancies. Particularly, the RF detects a very small number of the ACVS Cepheids, Delta Scuti, and Chemically Peculiar stars.  Also, the RF finds a gross overabundance of Double Mode RR Lyrae and Wolf-Rayet stars.  These artifacts result from sample selection bias. \label{fig:rfasaspred}
}
\end{figure*}

By training a Random Forest classifier on the \hipp and OGLE data as in \citet{2011rich} and applying that classification model to predict the class label of each object in ACVS, we obtain a 65.5\% correspondence with the ACVS labels for the 24\% of objects that have a confident ACVS label.  A table showing the correspondence of our predicted Random Forest classification labels with those of ACVS is plotted in Figure \ref{fig:rfasaspred}.  The Random Forest algorithm classifies 90\% and 79\% of the Mira and RR Lyrae, FM stars identified by ACVS, but shows much lower correspondence for other classes, such as Delta Scuti, Population II Cepheid, and RR Lyrae, FO.  Note that the Random Forest class taxonomy is finer than that used by ACVS, including twice as many classes; as such, the Random Forest has the ability to identify objects of rarer classes, such as T Tauri and Gamma Doradus stars.

However, there are serious problems that arise by running the analysis in this manner and ignoring the significant sample selection bias between the training and testing sets.  In Figure \ref{fig:traintestoffset} we saw that  the distribution of the training set of \hipp and OGLE sources is wildly different than the distribution of ASAS sources;
notably, regions of long-period, amplitude $<1$ sources and regions of short-period, high-amplitude sources are densely populated in ASAS but contain little or no training data.  As a consequence, a large proportion of the ASAS data set has no counterpart in the training set that closely matches its feature vector, meaning that it will likely be incorrectly identified by the Random Forest classifier as belonging to a physically different class of variable star.  One telling statistic is that for only 14.6\% of the ASAS objects does the Random Forest produce a posterior class probability of $\ge 0.5$, meaning that the classifier is only confident on the class predictions for 15\% of the entire ASAS ACVS catalog.

In Figure \ref{fig:rfasaspred} we find that many ASAS sources (9109 of 50,124, or 18.2\%) are  identified by the Random Forest classifier as being of RR Lyrae, DM type, a relatively rare type of doubly pulsating variable star.  This is far too many RR Lyrae, DM candidates; for comparison, \citet{2011AcA....61....1S} find only 91 RR Lyrae, DM candidates in the entire OGLE-III catalog, out of 16,836 total RR Lyrae candidates (0.5\%).  This artifact in our classification occurs because the RR Lyrae, DM objects have multiple pulsational modes, causing their data to poorly fold around a single period.  Because ASAS photometry is less precise than that of \hipp or OGLE, its folded light curves are considerably more noisy.  Consequently, for a large subset of ASAS sources that do not resemble any of the training data, the classifier's ``best guess" is RR Lyrae, DM because training light curves of that class most resemble ASAS data.  In addition, an artificially high number of Wolf Rayet and Beta Lyrae stars are found by the RF.  This deficiency of the off-the-shelf classifier illustrates the need for other  approaches.

\section{Methods to Treat Sample Selection Bias}
\label{sec:methods}

Above, sample selection bias was defined, its presence in astrophysical problems motivated, and its adverse effects exemplified with an example in variable star classification.  In this section, we will introduce three different principled approaches of treating sample selection bias, and argue that active learning is the most appropriate of these methods for dealing with astronomical sample biases.  Later, these methods will be compared using variable star data from the OGLE and \hipp missions.

\subsection{Importance Weighting}
\label{ss:impwt}
Under sample selection bias, standard generalization error estimation procedures, such as cross-validation, are biased, resulting in poor model selection for supervised methods.  To remedy this, importance weighting (IW) cross-validation is often used (see \citealt{sugi2005}, \citealt{huan2007}, and \citealt{sugi2007}).  Under this approach, the training examples are weighted by an empirical estimate of the ratio of test-to-training-set feature densities during the training procedure.  Specifically, when evaluating the statistical risk of the statistical model over the training data, the weights
\begin{equation}
\label{eq:iw}
w_i = \frac{\Prob_{\rm Test}(\x_i,y_i)}{\Prob_{\rm Train}(\x_i,y_i)} = \frac{\Prob_{\rm Test}(\x_i)\Prob_{\rm Test}(y_i|\x_i)}{\Prob_{\rm Train}(\x_i)\Prob_{\rm Train}(y_i|\x_i)} = \frac{\Prob_{\rm Test}(\x_i)}{\Prob_{\rm Train}(\x_i)}
\end{equation}
are typically used, where $\x_i$ is the feature vector and $y_i$ is the response variable (i.e., class) for training object $i$.  To achieve the last inequality in Equation \ref{eq:iw}, it is assumed that $\Prob_{\rm Test}(y_i|\x_i)=\Prob_{\rm Train}(y_i|\x_i)$, i.e. that the probability of a specific response given a feature vector is the same for training and testing sets.  In practice, this equality will probably not hold for the types of astrophysical data sets we are interested in: though the mapping from features to response values may be the same for the training an testing sets, the prior distributions over the responses, $y$, are different, in general.  Even in this situation, use of the ratio of feature densities---though imperfect---may still be useful, and is more tractable than using the joint feature-response densities\footnote{Note that we could alternatively rewrite the joint density as $\Prob(y_i)\Prob(\x_i | y_i)$.   It is unlikely that $\Prob_{\rm Test}(\x_i | y_i) = \Prob_{\rm Train}(\x_i | y_i)$ in most practical situations; however, if this were to hold then the importance weights would simply reduce to the ratio of response priors.}.
 Even so, in practice the training and testing feature densities are difficult to estimate (and their ratio is even harder to estimate) because they reside in high-dimensional feature spaces.   To overcome this,  Eqn. \ref{eq:iw} can be estimated via distribution matching (\citealt{huan2007}) or by fitting a probabilistic classifier to the classification problem of training vs. testing set and employing the output probability estimates (\citealt{zadr2004}).


Using the weights defined in Eqn. \ref{eq:iw} when training a classifier induces an estimation procedure  that gives higher importance to training set objects in regions of feature space that are relatively under-sampled by the training data, with respect to the testing density.   This enforces a higher penalty for making errors in regions of feature space that are under-represented by the training set.  This is sensible because, since the ultimate goal is to apply the model to predict the response of the testing data, we should attempt to do well at modeling the output in regions of feature space densely populated by testing data (and conversely ignore modeling regions devoid of testing data).  For the ASAS example, importance weighting will give large weights to the training data in the region of Amplitude $< 0.5$ and Period $> 100$ and affix small weights to data in the high-amplitude clump centered around a 300-day period.

Though this approach is useful in some problems, importance weighting  has been shown to be asymptotically sub-optimal when the statistical model is correctly specified\footnote{In other words, IW produces worse results than the analogous unweighted method if the parametric form of $\Prob(y|\x)$ is correct.} (\citealt{shimo2000}), and with flexible non-parametric models such as Random Forest we observe very little change in performance using IW (see \S\ref{sec:deb}).  An additional, more debilitating drawback is that IW requires  the support of the testing distribution be a \emph{subset} of the support of the training distribution\footnote{Else the weights, defined as the ratio of test-to-training set feature densities, explode, and the theoretical properties of the method no longer hold.}, which, in the types of supervised learning problems common in astrophysics, is rarely the case.

\subsection{Co-training}
\label{ss:cotrain}

In astronomical problems, we typically have much more unlabeled than labeled data.  This is due to both the pain-staking procedures by which labels must be accrued (e.g., by spectroscopic follow-up or manual assignment), and the fact that there are exponentially more dim, low signal-to-noise sources than bright, well-understood sources.  Recently, supervised classification algorithms have been developed that use both labeled and unlabeled examples to make decisions.  This class of models is referred to as \emph{semi-supervised} because learning is performed both on the instances with known response values and on the feature distribution of instances with no known response.  Semi-supervised methods such as \emph{co-training} and \emph{self-training} slowly augment the training set by iteratively adding the most confidently-classified test cases in the previous iteration.

Co-training was formalized by \citet{blum1998} as a method of building a classifier from scarce training data.  In this method, two separate classifiers, $h_1$ and $h_2$, are built on different (disjoint) sets of features, $\x_1$ and $\x_2$.  In an iteration, each classifier adds its $p$ most confidently labeled test instances to the training set of the \emph{other} classifier.  This process continues either for $N$ iterations or until all test data belong to the training set of both classifiers.  The final class predictions are determined by multiplying the class probabilities of each classifier, i.e. $p(y|\x) = h_1(y|\x_1)h_2(y|\x_2)$.  Co-training has shown impressive performance in situations where very few training examples are used to classify many test cases.  \citet{blum1998} use co-training in a two-class problem, using 12 labeled web pages to classify a corpus of 1051 unlabeled pages, achieving a 5\% error rate.

In the original co-training formulation, it was assumed that each object could be described by two different `views' (i.e. feature sets) of the data that were both redundant (each view of the object gives similar information) and conditionally independent given the true class label.  While this natural redundancy may be present in web page classification (e.g., the words on the web page and the words on pages linked to that page), it is not generally the case.  Later papers by \citet{goldman2000} and \citet{nigam2000} argue that even when a natural feature division does not exist, arbitrary or random feature splits produce better results than self-training (\citealt{nigam2000}), where a single classifier is built on all of the features whereby the most confidently classified testing instances are iteratively moved to the training set.

In the variable star classification paper of \citet{2009A&A...506..519D}, something akin to a single iteration of self-training was performed for CoRoT classification using OGLE training data, where candidate lists obtained with the first version of the classifier were used to select very probable class members amongst the testing set data for subsequent inclusion in the training set.  This augmentation procedure led to inclusion of an extra 114 sources into the training set. 

Both co-training and self-training are reasonable approaches to problems that suffer from sample selection bias because they iteratively move testing data to the training set, thereby gradually decreasing the amount of bias that exists between the two sets.  However, in any one step of the algorithm, only those data in a close neighborhood to existing training data will be confidently classified and made available to be moved to the training set.  Thus, as the iterations proceed, the dominant classes in the training data  diffuse into larger regions of feature space, potentially gaining undue influence over the testing data.  In addition, co-training and self-training will never predict classes that are rare or unrepresented in the training data, even if they are prominent in the testing data.   In \S\ref{sec:deb} we  apply both self-training and co-training to variable star classification, finding that these methods perform poorly in terms of overall error rate, especially for classes that are under-sampled by the training data.


\subsection{Active Learning}
\label{ss:actlearn}

An important feature to supervised problems in astronomy is that we often have the ability to selectively follow up on objects to ascertain their true nature.  For example, for different problems this can be achieved by targeted spectroscopic study, visualization of (folded) light curves, or querying of other databases and catalogs.  Consider astronomical source classification: while it is impractical to manually label all hundred-million plus objects that will be observed by Gaia and LSST, manual labeling of a small, judiciously chosen set of objects can greatly improve the accuracy of an automated supervised classifier.  This is the approach of \emph{active learning} (and in particular, pool-based active learning, \citealt{lewis1994}).   Under  pool-based AL for classification, an algorithm iteratively selects, out of the entire set of unlabeled data,  the object (or set of objects) that would give the expected maximal performance gains of the classification model, if its true label(s) were known.  The algorithm then queries the user to manually ascertain the science class of the object(s), whereby the supervised learner incorporates this information its subsequent training sets to improve upon the original classifier.  For a thorough review of active learning, see \citet{sett2009}.

Active learning has enjoyed wide use in  machine learning, with impressive results in many areas of application, such as text classification, speech recognition, image and video classification, and medical imaging (\citealt{lewis1994,tong2001,tong2002,yan2003,liu2004,tur2005}).  Begin with a training set $\mathcal{L}$ and testing set $\mathcal{U}$.  On each active learning iteration, we manually find the class of the testing set source, $\x' \in \mathcal{U}$, whose inclusion into $\mathcal{L}$ would most improve the classifier's performance on the testing data (according to some metric, see \S\ref{ss:query}).  These queried active learning samples tend to be data that reside in relatively dense regions of testing set feature space, $\Prob_{\rm Test}(\x)$, scarcely populated regions of training set feature space, $\Prob_{\rm Train}(\x)$, and in regions where the class identity is uncertain. 

For an appropriate selection metric, a small number of active learning samples will suffice in making the labeled set feature distribution resemble the unlabeled set distribution.  This approach is similar to the importance sampling approach of \citet{zadr2004}, who show that  if training set sources are resampled with respect to the appropriate (weighted) distribution, then the statistical risk of the classifier built on that data will minimize the statistical risk evaluated over all of the data.  The drawback to that approach is that it needs a relatively large initial training sample and requires that for all non-zero regions of $\Prob_{\rm Test}$, $\Prob_{\rm Train}$ also be non-zero.  On the other hand, the active learning approach to sample selection bias is to \emph{expand} the training set in a way that makes it most closely resemble the testing set, and thus these problems are avoided.


\subsubsection{Active Learning Query Function}
\label{ss:query}

Several  strategies have been proposed to determine which testing data about which active learning will query the ``human annotator."  Most of these prescriptions attempt to select data whose label, if known, would maximally help the classifier.  The simplest form of querying is \emph{uncertainty sampling} (\citealt{lewis1994}), by which on each iteration, the training datum with highest label uncertainty (measured, e.g., by entropy or margin) is queried for manual identification.  Though simple, this approach does not explicitly consider changes to the overall error rate of the classifier, and is prone to select outlying points that have little influence in the classification of the other testing data.  

Since we have an explicit goal of minimizing the classification error rate over the entire set of testing data, it is sensible to consider this metric explicitly when queuing data for AL.  This is the approach taken by the \emph{expected error reduction} strategies (\citealt{roy2001toward}), where on each iteration the algorithm queries the testing point whose inclusion into the training set would produce the smallest classification error rate (statistical risk) over the testing set.  These methods operate by iteratively adding each testing point to the training set and retraining the classifier\footnote{For many machine learning algorithms, fast incremental updating algorithms exist, making this approach tractable.}.  However, because the true labels of the training data are not known \emph{a priori}, one must also iterate over the possible labels of the training data, and can only compute an estimate of the expected decrease in testing error rate by approximating the error under all possible labels of all testing data.  For common astronomy data sets, with $\gtrsim 10^5$ objects, expected error reduction is impractical.  A viable alternative is \emph{variance reduction} (\citealt{cohn1996}), where the testing object that minimizes the classifier's variance is selected on each iteration.  Since a classifier's error can be decomposed into variance plus squared-bias plus label noise\footnote{Classifier variance measures the variability in a classifier with respect to the actual training set used, classifier bias is the amount of discrepancy between the true labels and the expected prediction of a classifier (averaged over all possible training sets), and label noise is the amount of error in the training set labels.}, minimizing the variance amounts to minimizing the error rate; also, for many models, the variance can be written in closed form, circumventing any costly computations.

In this paper, we consider two different selection criteria.  The first criterion is motivated by importance weighting and the second is motivated by selecting the sources whose inclusion into the training set would produce the most total change in the predicted class probabilities for the testing sources.  To meet these criteria, we revisit the Random Forest classifier.  For each of $B$ bootstrap samples from the training set, we build a decision tree, $\theta_b$, which predicts the class of each object from its feature vector, $\x$.  The Random Forest class probability of class $y$ is simply the empirical proportion,
\begin{equation}
\label{eq:rfprob}
\widehat{P}_{{\rm RF}} (y|\x) = \frac{1}{B} \sum_{b=1}^B \theta_{b}(y|\x)
\end{equation}
of the $B$ trees that predict class $y$.  Additionally, the Random Forest provides a measure of the \emph{proximity} of any two feature vectors with respect to the ensemble of decision trees, defined as
\begin{equation}
\rho(\x',\x) = \frac{1}{B} \sum_{b=1}^B I(\x \in T_b(\x'))
\end{equation}
which  is the proportion of trees for which the two objects $\x$ and $\x'$ fall in the same terminal node, where  $I(\cdot)$ is a boolean indicator function.  Here, we use the notation $T_b(\x')$ to denote the terminal node of feature vector $\x'$ in tree $b$.

Heuristically, sample selection bias causes problems in the building of a classifier principally because large density regions of testing data are not well represented by the training data. Our first AL selection procedure uses this heuristic argument to select the testing point, $\x' \in \mathcal{U}$, whose feature density is most under-sampled by the training data, as measured by the ratio of the two densities.  
This amounts to choosing AL samples that maximize
\begin{equation}
\label{eq:alrf1}
S_1(\x') = \frac{\Prob_{\rm Test}(\x')}{\Prob_{\rm Train}(\x')} \approx \frac{\sum_{\x \in \mathcal{U}} \rho(\x',\x)/N_{\rm Test}}{\sum_{\z \in \mathcal{L}} \rho(\x',\z)/N_{\rm Train}}
\end{equation}
where we estimate the training and testing set densities at $\x'$ by averaging the RF proximity measure over the set of training ($\mathcal{L}$) and testing ($\mathcal{U}$) sets, respectively.  The expression $\sum_{\x \in \mathcal{U}}\rho(\x',\x)/N_{\rm Test}$, is the average, over the trees in the forest, of the proportion of testing data with which $\x'$ shares a terminal node.  The estimate of the probability density at $\x'$ would need be normalized by the average volume of the terminal nodes of $\x'$; however, since Equation \ref{eq:alrf1}  considers the ratio of two such densities at $\x'$, the average volume terms cancel, giving the above expression.

 Our second AL  selection criterion is to choose the testing example, $\x' \in \mathcal{U}$, that maximizes the total amount of change in the predicted probabilities for the testing data.  This is a reasonable metric because it says that we will only spend time manually annotating the testing data whose labels  most affect the predicted classifications.  To achieve this, we create a selection metric that attempts to choose the $\x'$ that maximizes the total change, summed over the testing set, of the  RF probability vectors (as measured using the $\ell_1$ norm).  An approximate solution to this problem is to choose the testing data points that maximize
\begin{equation}
\label{eq:alrf2}
S_2(\x') = \frac{\sum_{\x \in \mathcal{U}}\rho(\x',\x)(1 - \max_y \widehat{P}_{{\rm RF}}(y|\x))}{\sum_{\z \in \mathcal{L}} \rho(\x',\z) +1}
\end{equation}
where the Random Forest  probability, $\widehat{P}_{{\rm RF}}(y|\x)$, is defined in Equation \ref{eq:rfprob}.
In Appendix \ref{app:alrf} we work out the details of deriving Eqn. (\ref{eq:alrf2})  from the stated goal of selecting testing points whose labels maximally affect the total change of the Random Forest predicted probabilities over $\mathcal{U}$.	 

The key elements to Eqns. \ref{eq:alrf1}--\ref{eq:alrf2} are (1) the testing set density, represented by $\sum_{\x \in \mathcal{U}} \rho(\x',\x)$ is in the numerator, and (2) the training set density, represented by $\sum_{\z \in \mathcal{L}} \rho(\x',\z)$, is in the denominator.  This means that we will choose instances that are in close proximity to many testing points and are far from any training data, thereby reducing sample selection bias.  In addition, $S_2$ is a weighted version of $S_1$ with the Random Forest prediction uncertainty, represented by $1 - \max_y \widehat{P}_{{\rm RF}}(y|\x)$, in the numerator.  This means that $S_2$ gives higher weight to those testing points that are difficult to classify thereby causing the algorithm to focus more attention along class boundaries, which should lead to better performance of the classifier.

\subsubsection{Batch-Mode Active Learning}

In typical active learning applications, queries are chosen in serial.  However, in most astronomical applications, it makes more sense to query several testing set objects at once, in \emph{batch mode}; for instance, in a typical observing run multiple objects are queued for follow-up observation.  In the variable star classification problem, we determine that the best use of users' time is to supply them with dozens of sources to label at one sitting.

The challenge with batch-mode AL is to determine how to best choose multiple testing instances at once.  Selecting the top few candidates is typically suboptimal because those objects generally lie in the same region of feature space, as is obvious from analyzing the criteria in Eqns. \ref{eq:alrf1}-\ref{eq:alrf2}.  Heuristic methods have been devised that create diversity in the batch of AL samples for a particular classifier (e.g., \citealt{brinker2003} for SVMs).  In our use of AL, we sample batches of AL samples by treating the criterion function as a probability sampling density, i.e., $\Prob(\textrm{select } \x') \propto S_1(\x')$.  In \S\ref{sec:deb} we compare this density method, which we call AL-d, to a method that selects the top candidates on each AL iteration, which we refer to as AL-t.

\subsubsection{Crowdsourcing Labels}
\label{ss:crowdsource}

Most active learning papers assume that labels can be found, without noise, for any queried data point.  In typical astronomical applications, this will not be the case.  For instance, after follow-up observations of an object, its true nature might still be difficult to ascertain and will often  remain unknown.  Indeed, in classifying variable stars, users will sometimes have difficulty in obtaining the true class of an object, especially for noisy or aperiodic sources.  This causes two complications in the AL process:
\begin{enumerate}
\item some queried sources will still have an unknown label after manual classification, and
\item a few sources will be annotated with an \emph{incorrect} label.
\end{enumerate}
The first difficulty means that we expect to receive user labels for only a fraction of the queried sources; to avoid wasting costly user time, we attempt to select AL sources that users will have a higher probability of successfully labeling (in \S \ref{ss:cost} we describe how this is achieved by using a cost function).  To overcome the second complication, we use crowdsourcing, where several users are presented with the same set of AL sources.  The idea behind crowdsourcing is that by using the combined set of information about each object from multiple users, we are able to suppress the noise in the manual labeling process.

A difficulty in crowdsourcing is in simultaneously predicting the best label and judging the accuracy of each annotator from a set of user responses.   Users are likely to disagree on some objects, so determining a true label can often be tricky.  However, because each annotator has a different skill level, we should give more credence to the labels of the more adept users in deciding on a  label.  In the active learning paper of \citet{donmez2009}, a novel, yet simple method called {\tt IEThresh} was introduced to filter out the less-adept users in crowdsourcing labels.  Their basic approach is to start each user with the same prior skill level.  Then, as the AL iterations progress, users whose responses agree with the consensus votes of the crowd are given higher `reward'.  The skill level of each user is determined by the upper confidence interval (UI) of the mean reward of all their previous labels.  For each subsequent iteration, only those users whose UIs are higher than $\epsilon$ times the UI of the best annotator are included in the vote for the class of that object.  Even if a particular user's label is not used in a vote, their reward level can change, meaning that users are able to drift in and out of the decision-making process over time.  

In \S\ref{sec:results}, we use the {\tt IEThresh} algorithm with $\epsilon=0.85$ to crowdsource the ASAS labels.  In addition, for a source to be included in the training set, we require that at least 70\% of users who looked at the source return a label.  This strict policy is implemented so that only the most confident AL sources are moved to the training set so as to avoid including incorrectly labeled objects.

\subsubsection{Cost of Manual Labeling}
\label{ss:cost}

Standard active learning methods assume that the cost of attaining a label is the same for every data point, and thus aim to minimize the total number of queries performed (or equivalently achieve the lowest error rate for a given number of queries).  This assumption is not valid for variable star classification problem, for a variety of reasons.  First, higher signal-to-noise light curves with larger number of epochs will be, on average, easier to manually label than sparser, noisier light curves.  Second, a star that has been observed and cataloged by multiple surveys (for instance, it is in the SDSS footprint) will have more archival data with which to determine its true class.  Third, depending on its coordinates, a star may or may not be readily available for spectral follow-up.  To avoid wasting user time on impossible-to-classify objects, these factors must be taken into account when choosing AL samples.

In applying AL to variable star classification, we treat the cost as a multiplicative factor on the querying criteria.  That is, the AL function is $S(\x') = S_1(\x')(1-C(\x'))$, where the cost function, $C(\x')$, is
\begin{equation}
C(\x') = \Prob(\x' \textrm{ cannot be manually labeled }  |  \x' \textrm{ is queried}),
\end{equation}
i.e., the cost function is the probability that a user (or set of users) cannot actually determine a label for that source, given that the user was given that object to manually study\footnote{Other definitions of the cost are possible, such as the time necessary for a user to manually label a source or the user disagreement rate.  As formulated, our ``cost" function measures the inutility of the user on each particular source.}.  High cost means that we will avoid querying that object.  Inclusion of a cost function deters us from wasting valuable user time on objects that are too noisy or sparsely sampled to determine their science class.  In \S\ref{sec:results} we describe how we model the cost and derive an empirical cost estimate for each object in the ASAS testing set.

\subsubsection{Stopping Criterion}

Insofar as the aim of active learning is to improve the performance of a classifier to the greatest extent possible with as little effort as possible, we must determine when to stop manually labeling sources.  A reasonable rule of thumb is to stop querying data for active learning when the effort needed to acquire the new labels is larger than the benefit that those labels  have on the classifier's performance.  However, it is often difficult to compare these gains and losses, especially for problems where there do not exist ground truths with which to judge the classifier performance nor good metrics to measure gains and losses.  Alternatively, one can track the intrinsic stability of the classifier (e.g., by measuring its average confidence over the testing set), and stop when a plateau is reached (cf. \citealt{vlachos2008,olsson2009}).  In our implementation of AL, we choose to run iterations until the performance of the classifier levels off (as judged by a few intrinsic and extrinsic metrics, see \S\ref{sec:results}).


\section{Experiment: OGLE and \hipp Variable Stars}
\label{sec:deb}

\begin{figure}
\begin{center}
\includegraphics[angle=0,width=6.0in]{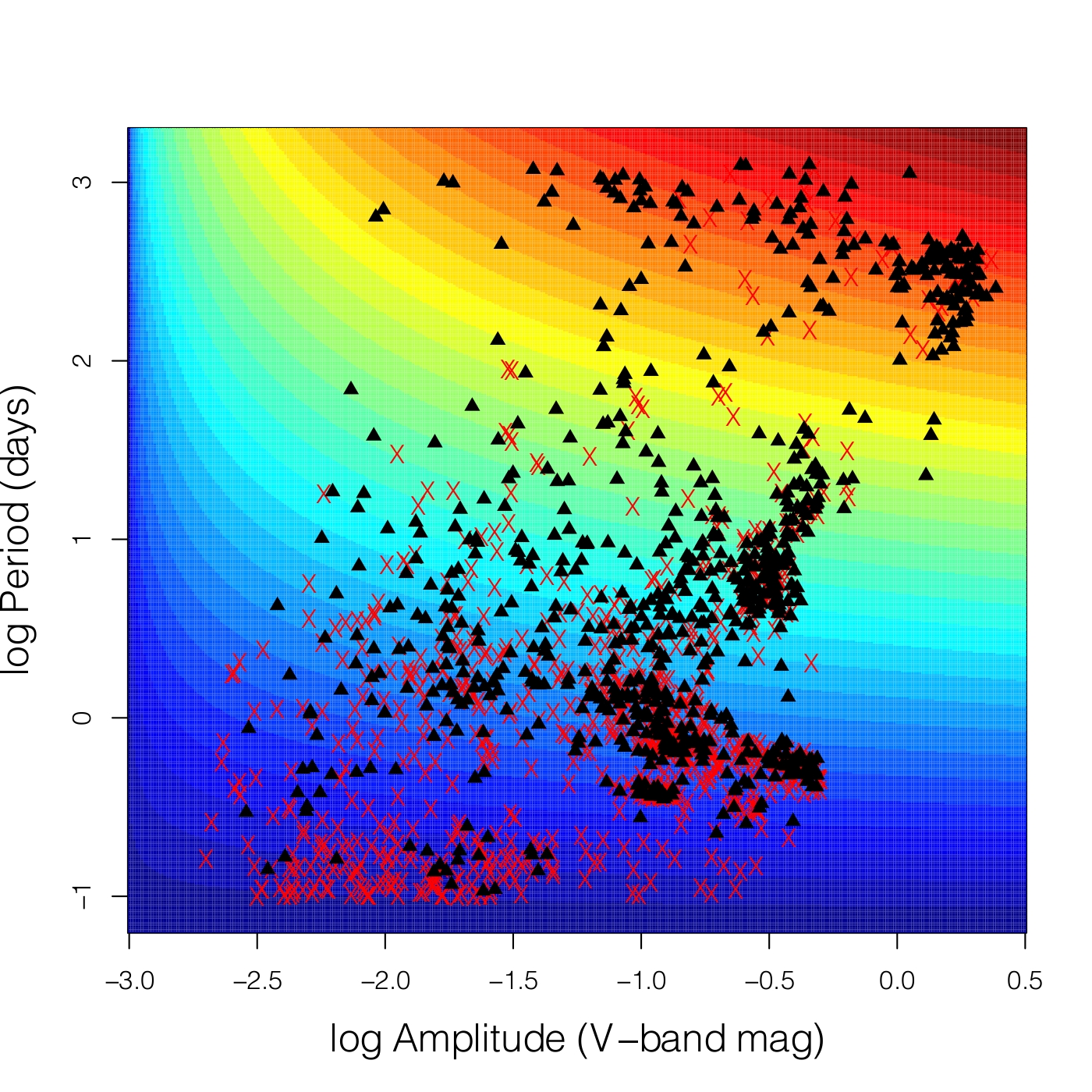}
\end{center}
\caption{Training (black $\blacktriangle$) and testing (red x) data for the simulated example using OGLE \& \hipp data.  The 771 training data were randomly sampled from the original 1542 sources according to the sampling distribution plotted in color.  Using this sampling scheme, we create sample selection bias by over-sampling long-period, high-amplitude stars and under-sampling the short-period, low-amplitude sources.  \label{fig:debtrte} }
\end{figure}

In this section, we test the effectiveness of the various methods proposed in \S\ref{sec:methods} in combating sample selection bias for variable star classification.  Starting with the set of 1542 well-understood, confidently-labeled variable stars from \citet{2007debo}, we randomly draw a sample of 721 training sources according to a selection function, $\Gamma$, that varies across the amplitude-period plane as
\begin{equation}
\Gamma(\x) \propto \log(\textrm{period }\x) \cdot \log(\textrm{amplitude }\x)^{1/4}.
\end{equation}
This selection function is devised so that the training set under-samples short-period, small-amplitude variable stars.  The resultant training and testing sets are plotted in the amplitude-period plane, along with the training set selection function, in Figure \ref{fig:debtrte}.  

Distributional mismatch between the training and testing sets causes an off-the-shelf Random Forest classifier to perform poorly for short-period small-amplitude sources.  The median overall error rate for a Random Forest classifier trained on the training data and applied to classify the testing data is 29.1\%.  This is 32\% larger than the 10-fold cross-validation error rate of 21.8\% on the entire set of 1542 sources (see \citealt{2011rich}; the error rate quoted here is slightly lower due to the addition of new features).  The average error rate for testing set objects with period smaller than 0.5 days is 36.1\%.

  To treat the sample selection bias, we use each of the following methods:
\begin{itemize}
\item Importance weighting.  A single Random Forest is built on the training set, with class-wise\footnote{In importance weighting, ratios of feature densities are typically used as the weights.  However, in our implementation of Random Forest, weights may only be defined by class.} importance sampling weights defined as the ratio of the testing set to training set class proportions\footnote{Since we know the true class of each object, we are able to use this information to derive the weights.  In a real problem, the feature or class densities would need to be estimated.}.
\item Self-training and co-training.  Each algorithm is repeated for 100 iterations, where on each iteration the most confident 3 testing set objects are added to the training set.  For co-training, we use both random feature splits (CT) and periodic versus non-periodic features (CT.p).
\item Active learning.  Using the metrics in Equations \ref{eq:alrf1} (AL1) and \ref{eq:alrf2} (AL2), we perform 10 rounds of active learning, with batch size of 10 objects selected on each round.  The classifier is retrained on the available labeled data after each round.  Testing set objects are selected for manual labeling either by treating the selection metrics as probability distributions (AL1.d, AL2.d), or by taking the top candidates (AL1.t, AL2.t).  We also compare to an AL method that selects objects completely at random (AL.rand). 
\end{itemize}
For each of the active learning approaches, we evaluate the error rate only over those testing set objects that are not queried by the algorithm.  This way we do not artificially decrease the error rate by evaluating sources whose labels have been manually obtained.  Note that for this experiment, we have assumed that the true labels can be manually obtained with no error. 

\begin{figure}
\begin{center}
\includegraphics[angle=0,width=6.5in]{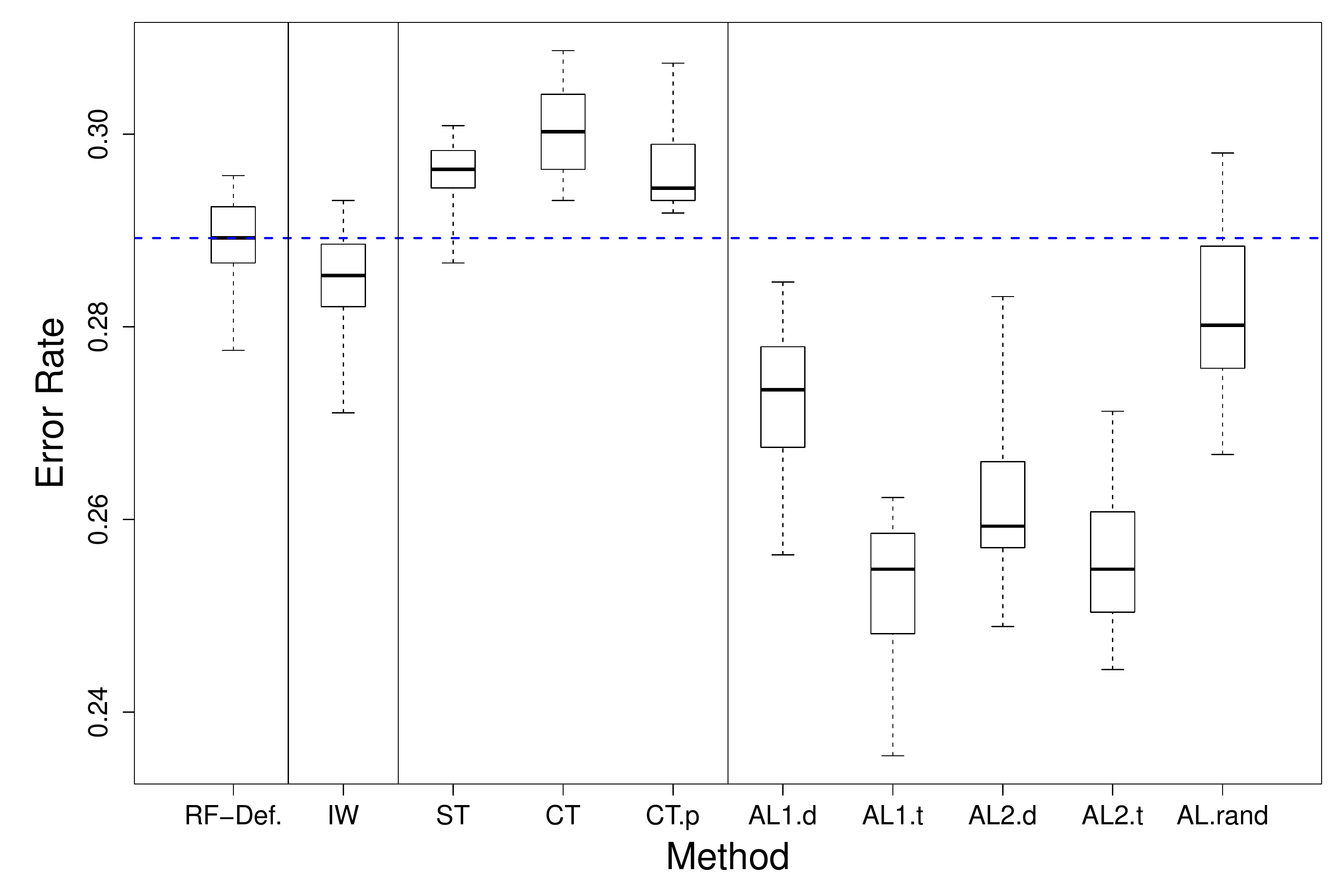}
\end{center}
\caption{Error rates, evaluated over the testing set, of 10 different methods applied to the OGLE \& \hipp simulated data set of 771 training and 771 testing samples.  Due to sample selection bias, the default Random Forest (RF-Def.) is ineffective.  Importance weighting (IW) improves upon the RF only slightly.  The co-training and self-training methods produce an increased error rate.  Only the active learning approaches yield any significant gains in the performance of the classifier over the testing set.  Note that the AL methods were evaluated over those testing data not in the active learning sample.  No large difference is found between the two AL metrics, but both outperform the random selection of AL samples.  Note that each boxplot displays the 25th and 75th quantiles as the edges of the boxes, with the center line denoting the median and the whiskers extending to the minimum and maximum. \label{fig:deberr} }
\end{figure}

Distributions of the classification error rates for each method, obtained over 20 repetitions, are plotted in Figure \ref{fig:deberr}.  The largest improvement in error rate is obtained by both AL1.t and AL2.t (25.5\% error rate), followed by AL2.d (25.9\%).  Quoted results for the active learning methods are after querying 100 training set objects (10 AL batches of size 10).  AL1.d lags well behind the performance of these other AL querying functions.  None of the other methods produces a significant decrease in the error rate of the classifier.  Indeed, the ST and CT approaches cause an \emph{increase} in the overall error rate.  IW produces a slight decrease in the error rate, by an average of 0.4\%, which represents 3 correct classifications.  An important observation is that the AL.rand approach of randomly queuing observations for manual labeling does not perform well  compared to the more principled AL approaches.

\begin{figure}
\begin{center}
\includegraphics[angle=0,width=6.5in]{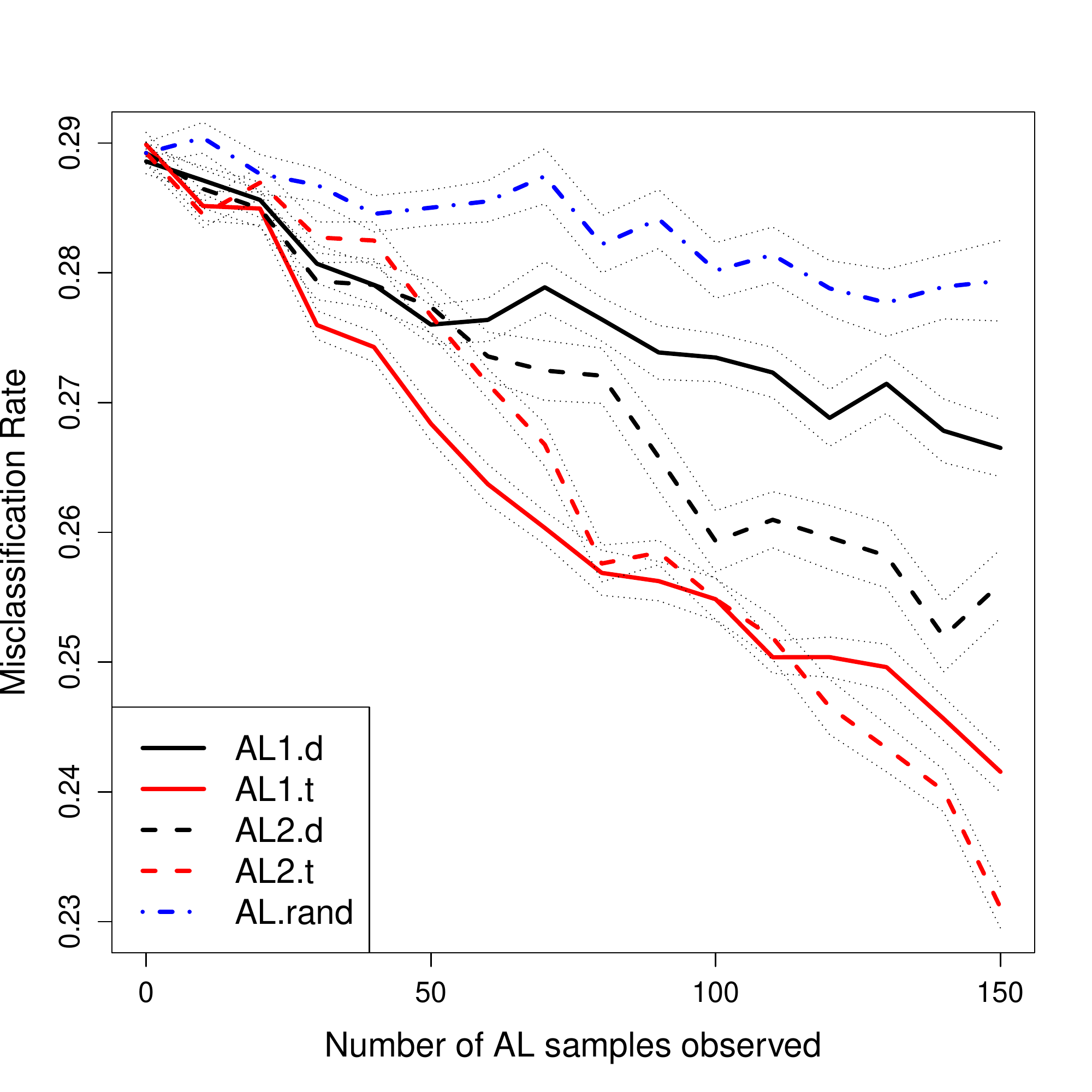}
\end{center}
\caption{Performance of the active learning approaches for the OGLE \& \hipp classification experiment.  Both AL1 and AL2 dominate the performance of AL.rand, but there is no clear winner between these two approaches.  AL1.t performs best for the first few iterations, but is overtaken by AL2.t after 100 samples are queried.  AL2.d performs significantly better than AL1.d after about 50 iterations.  For each method, the mean error rate---evaluated over the testing set not included in the AL sample---is plotted along with $\pm 1$ standard error bands.  \label{fig:debal} }
\end{figure}

Figure \ref{fig:debal} depicts the error rate of the AL approaches as a function of the total number of objects queried.  Between the AL1 and AL2 metrics, there is no clear winner, but once large numbers of samples have been observed AL2.d and AL2.t perform better than their AL1 counterparts.  We also find in Figure \ref{fig:debal} that the AL.d approaches---where objects are drawn with probability proportional to the AL criterion---perform worse than the approaches that always select the top AL candidates.  This is unexpected, as selecting only the top methods in batch mode produces samples of objects from the same region in feature space, causing an inefficient use of follow-up resources.  However, this observed better performance by the AL.t strategies may be an artifact of using small batch sizes (10 objects); in the application of active learning to ASAS, we typically use batch sizes $>50$.

Active learning is able to significantly improve the classification error rate on the set of OGLE \& \hipp testing data because it selectively probes regions of feature space where class labels, if known, would most influence the classifications of a large number of testing data.  For the OGLE and \hipp variable star data, sets of low-amplitude, short-period stars are selected by the AL algorithm, which in turn improve the error rates within the science classes populated by these types of stars, without increasing error rates within the classes that are highly sampled by the training set.  We make this more concrete in Table \ref{tab:deb}, where the classifier error rates within a few select classes are shown.  The active learning classifiers show substantial improvement, on average, over the default Random Forest for the classes which are most under-sampled by the training data and show no increase in the error rates for the classes that are most over-represented in the training set.

\begin{deluxetable}{cccrrrrrrrrrr}
\tabletypesize{\footnotesize}
\tablewidth{7.0in}
\tablecolumns{13}
\tablecaption{Error rates, in \%, over all testing data, and for those testing data within selected science classes in the OGLE \& \hipp experiment.  The first set of classes are those most under-represented in the training data.  The second set are those most over-represented in the training data.  Several methods for sample selection bias reduction are compared. \label{tab:deb}}
\tablehead{   
 \colhead{Science Class} &
 \colhead{$N_{\rm Train}$} &
 \colhead{$N_{\rm Test}$} &
 \colhead{RF\tablenotemark{a}} &
 \colhead{IW } &
 \colhead{ST } &
 \colhead{CT } &
 \colhead{CT.p } &
 \colhead{AL1.d\tablenotemark{b}} &
 \colhead{AL1.t\tablenotemark{b}} &
 \colhead{AL2.d\tablenotemark{b}} &
 \colhead{AL2.t\tablenotemark{b}} &
 \colhead{AL.rand\tablenotemark{b}}  
  }
\startdata
\hline
All &771 &771 &28.9 &28.5 &29.6 &30.0 &29.4 & 27.3 & 25.5 & 25.9 & 25.5 &   28.0\\
\hline
Delta Scuti &  25 &89 &  15.7 &15.7 &15.7 &15.7 &14.6 & 15.4 & 14.0 & 15.6 & 21.3 &   12.3\\
Beta Cephei &9 &30 &    95.0 &91.7 &96.7 &96.7 &96.7 & 90.7 & 87.5 & 88.9 & 84.0 &   90.7\\
W Ursa Maj. &16 &43 &40.7 &36.0 &51.2 &60.5 &61.6  &27.0 & 27.3 & 27.1 & 19.2 &   30.1\\
\hline
Mira & 121 &23 & 8.7 & 8.7 & 8.7 & 8.7 & 4.3 &  9.1 &  8.7 &  8.7 &  8.7 &    9.8\\
Semi-Reg. PV &33 &9 & 33.3 &33.3 &33.3 &33.3 &33.3 & 33.3 & 33.3 & 33.3 & 33.3 &   35.4\\
 Class. Cepheid &122 &68 &  2.9 & 2.9 & 1.5 & 1.5 & 1.5 &  3.1 &  1.5 &  1.6 &  1.5 &    2.8
\enddata
\tablenotetext{a}{Default Random Forest.}
\tablenotetext{b}{Errors evaluated over all objects not in the active learning sample.}
\end{deluxetable}

\section{{\tt ALLSTARS}: Active Learning Light Curve Web Interface}
\label{sec:allstars}

\begin{figure}
\begin{center}
\includegraphics[angle=0,width=5.2in]{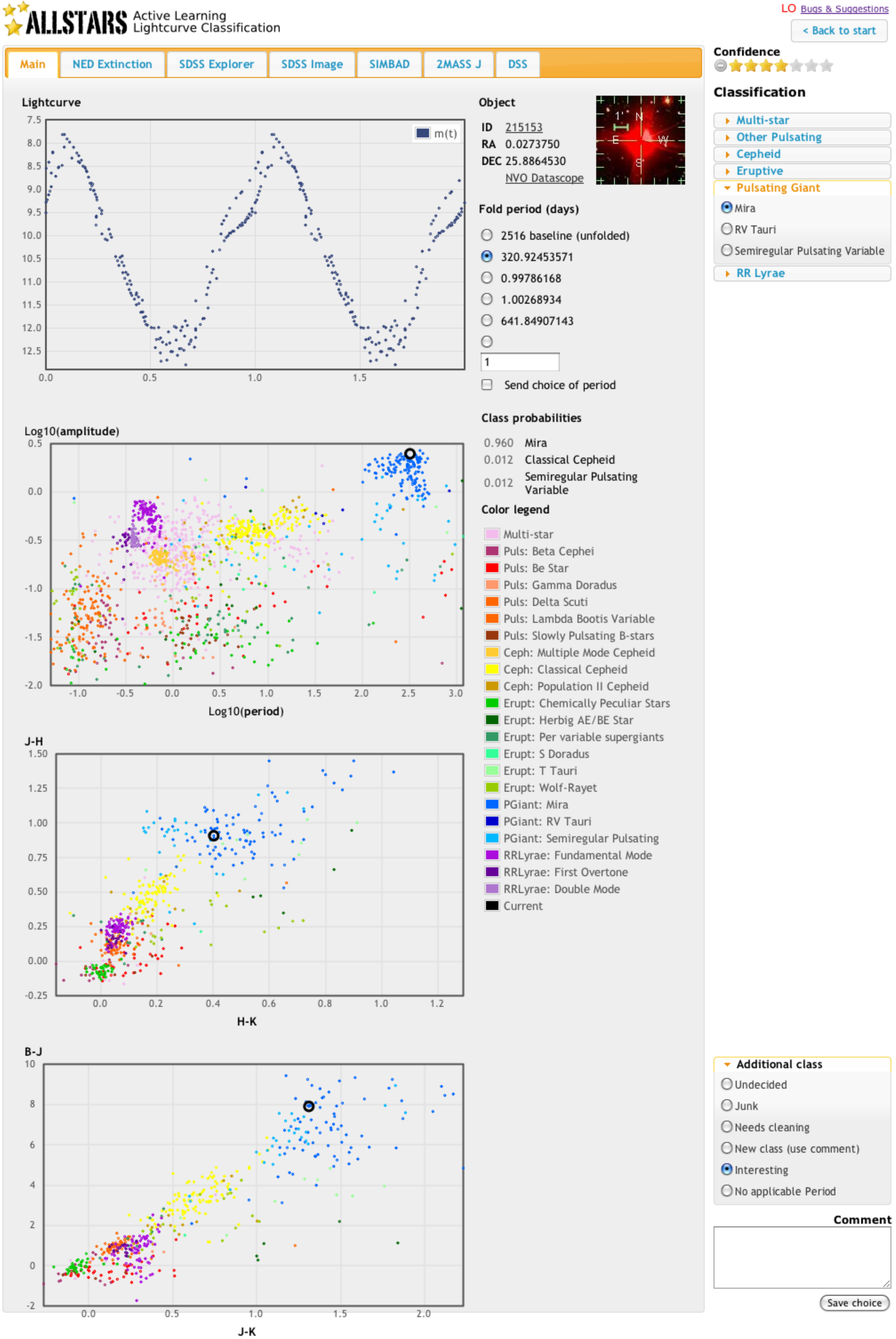}
\end{center}
\caption{ Screen shot of the {\tt ALLSTARS} web interface.  Here, a Mira variable from the ASAS survey has been queried by the user.  From top to bottom, the user is provided a (folded) ASAS light curve of the source, its location in amplitude-period space, its $J-H$ vs. $H-K$, and its $B-J$ vs. $J-K$ colors.  At the top of the page are several tabs which link to external resources.  On the left margin the user can make and submit a classification for the source.  \label{fig:allstars}}
\end{figure}

We developed the {\tt ALLSTARS} (Active Learning Lightcurve classification Service) web based tool as the crowdsourcing user interface to our active learning software.  For each active learning iteration, this website displays to a user the set of AL-queried sources.  For each source, users are given access to eight external web resources in addition to several feature space visualizations to facilitate manual classification of that source.  A screen shot of the {\tt ALLSTARS} web interface is in Figure \ref{fig:allstars}.  Additionally, for each source a user may make a science classification, a rating of their confidence, a data quality classification, can tag the source as interesting, and also may provide comments and store a manually-determined period.  This set of information is used to determine the class of each of the active learning queried sources and to decide which subset of those sources to add to the training set.  

{\tt ALLSTARS} was built using a combination of {\tt javascript}, {\tt PHP}, and {\tt Python} which accesses a {\tt MySQL} database.  Backend feature generators, active learning and classification algorithms were implemented using a combination of {\tt Python}, {\tt C} and {\tt R}.  The interactive plots are generated using the {\tt Flot jQuery}\footnote{{\tt Flot} is a  {\tt Javascript} plotting library downloadable from \url{http://code.google.com/p/flot/}.} package.  External resources made available for classifying each source are:
\begin{itemize}
\item NED Extinction Calculator:     \url{http://ned.ipac.caltech.edu/forms/calculator.html}
\item SDSS DR7 Explorer:              \url{http://cas.sdss.org/dr7/en/tools/explore/obj.asp}
\item SDSS DR7 Navigate Tool:         \url{http://cas.sdss.org/dr7/en/tools/chart/navi.asp}
\item SIMBAD Query by coordinates:    \url{http://simbad.u-strasbg.fr/simbad/sim-fcoo}
\item 2MASS Interactive Image ($J$-band):    \url{http://irsa.ipac.caltech.edu/applications/2MASS/IM/interactive.html}
\item SkyView Original DSS image:     \url{http://skyview.gsfc.nasa.gov/cgi-bin/query.pl}
\item NVO DataScope:                  \url{http://heasarc.gsfc.nasa.gov/cgi-bin/vo/datascope/init.pl}
\item DotAstro LightCurve Warehouse:  \url{http://dotastro.org/}
\end{itemize}

The initial page for a source includes two color-color plots: $B-J$ vs. $J-K$ and $J-H$ vs. $H-K$, using colors from the SIMBAD source which best matches the location of the given source.  The  source is also shown on a log-amplitude vs. log-period plot, with sources from the initial \hipp and OGLE training set displayed in the background.  These sources are discriminated using 21 different colors which represent most science classes to which the user may classify.  An interactive magnitude vs. time light curve plot is also shown, with options to display it either unfolded, folded on any of the three most significant periods, or folded using a user entered or zoom-box generated period.  The chosen period also updates a black circle on the amplitude-period plot.  Also available on this initial page are the top three algorithm classifications and their confidences. 

{\tt ALLSTARS} can be used to display any source available in the \url{http://dotAstro.org} Lightcurve Warehouse, allowing a registered user to make a science classification, assess data quality, note a manually found period, or add additional comments for that source.  This web interface is an extremely useful tool, not only for performing active learning for variable star classification, but also for following up on outliers discovered via unsupervised learning, for finding typical examples of light curves of desired science classes, and to manually search through subsets of the {\tt dotAstro} data warehouse.

\section{Application of Active Learning to classify ASAS Variable Stars}
\label{sec:results}

We use the active learning methodology presented in \S\ref{ss:actlearn} to classify all of ACVS (see \S\ref{ss:acvs}) starting with the combined \hipp and OGLE training set.  We employ the $S_2$ AL query function (Equation \ref{eq:alrf2}), treating it as a probability distribution (AL2.d in \S\ref{sec:deb}), and selecting 50 AL candidates on each of 9 iterations (except for the first iteration, where 75 AL candidates were chosen).  For a cost function, we employ data from our first AL iteration to train a logistic regression model to predict cost as a function of \verb freq_signif, the statistical significance of the estimated first frequency\footnote{This will bias us away from selecting aperiodic sources, such as T Tauri.  However, this is a reasonable approach because (1) there are simply too many aperiodic sources that are impossible to classify manually, and (2) in AL we draw a random sample from the $S_2(\x') * (1-C(\x'))$ meaning that we are still very likely to select some interesting aperiodic sources with high $S_2$ score.}.

\begin{figure}
\begin{center}
\includegraphics[angle=0,width=5.5in]{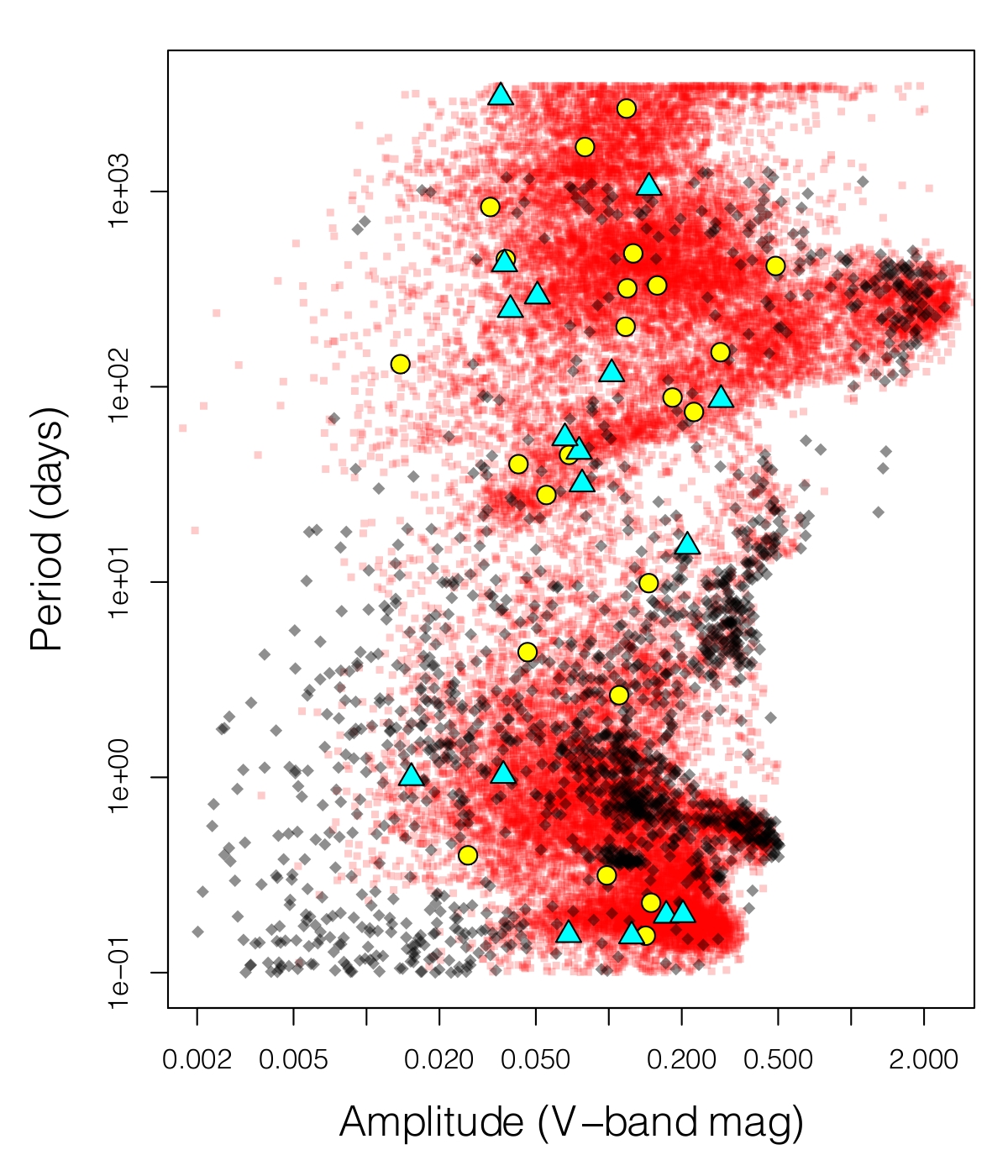}
\end{center}
\caption{ Active learning samples on a single iteration of the algorithm.  Yellow circles signify points that at least 65\% of users were able to classify.  These points are included on subsequent iterations of the algorithm.  Cyan triangles signify variable stars that were queried, but for which fewer than 65\% of users were able to classify.  Black diamonds and red squares are the original training and testing data, as in Figure \ref{fig:traintestoffset}. \label{fig:asasalsel}}
\end{figure}

A total of 11 users classified sources using the {\tt ALLSTARS} web interface.  To help train new users, the beginning of each iteration was populated with  14-18 high-confidence sources\footnote{As to not throw away useful annotations, these classifications were used along with the AL samples.}.  A total of 615 sources were observed by users (this represents 1.2\% of the ACVS catalog).  The average user classified 137 sources, with a range from 21--474.  User responses were combined using the crowdsourcing methodology in \S\ref{ss:crowdsource}.  This led to the inclusion of 415 ASAS sources (67\% of all sources that were studied manually) into the training set.  In Figure \ref{fig:asasalsel} we plot the AL queried data from one iteration in the amplitude-period plane, highlighting those which were selected for inclusion in the training set.

\begin{figure}
\begin{center}
\includegraphics[angle=0,width=6.5in]{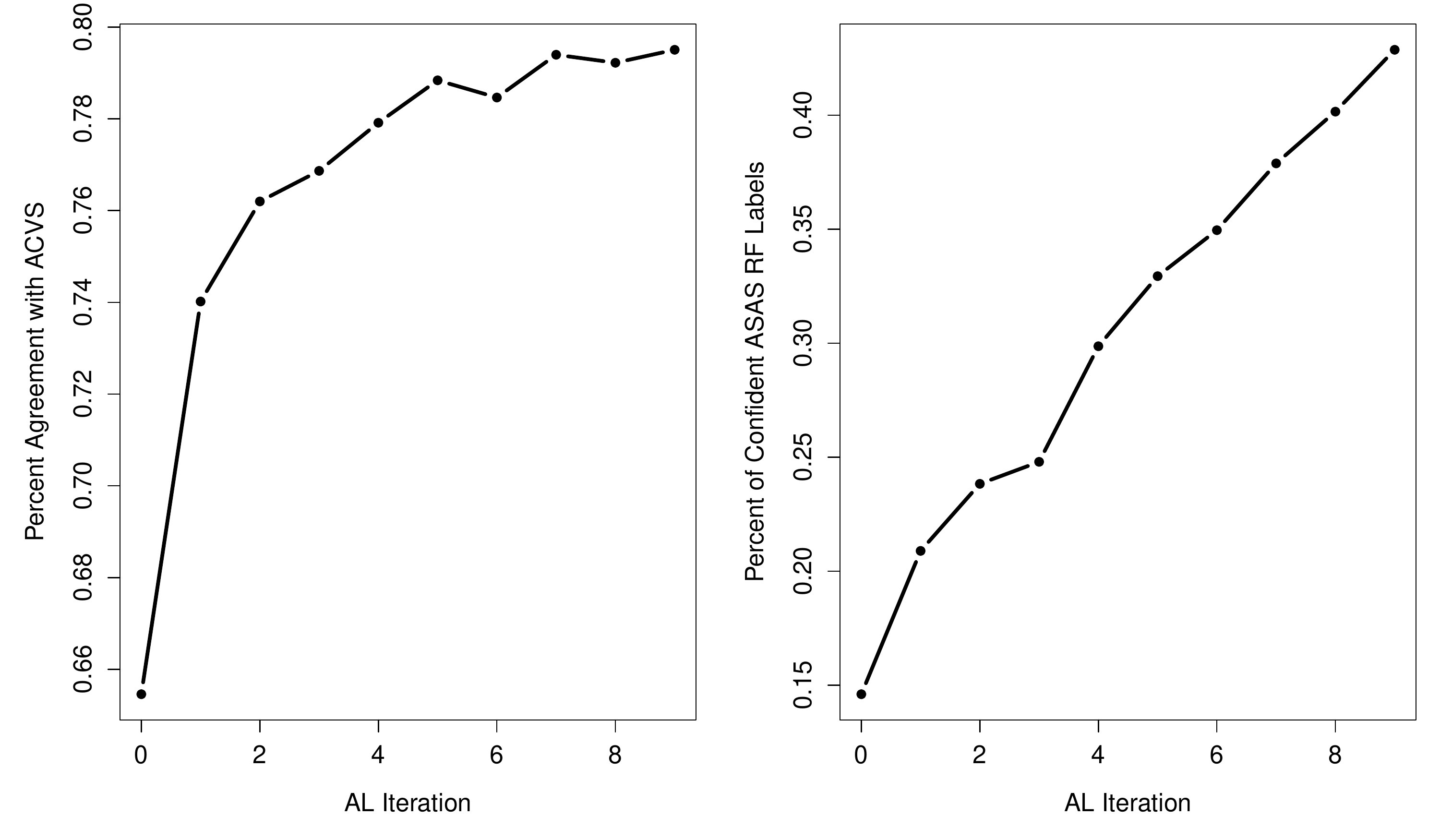}
\end{center}
\caption{ Left: Percent agreement of the Random Forest classifier with the ACVS labels, as a function of AL iteration.  Right: Percent of ASAS data with confident RF classification (posterior probability $>0.5$), as a function of AL iteration.  In the percent agreement with ACVS metric, performance increases dramatically in the first couple of iterations and then slowly levels off.  In the percent of confident RF labels, the performance increases steadily.   \label{fig:alrates}}
\end{figure}

\begin{figure}
\begin{center}
\includegraphics[angle=0,width=6.5in]{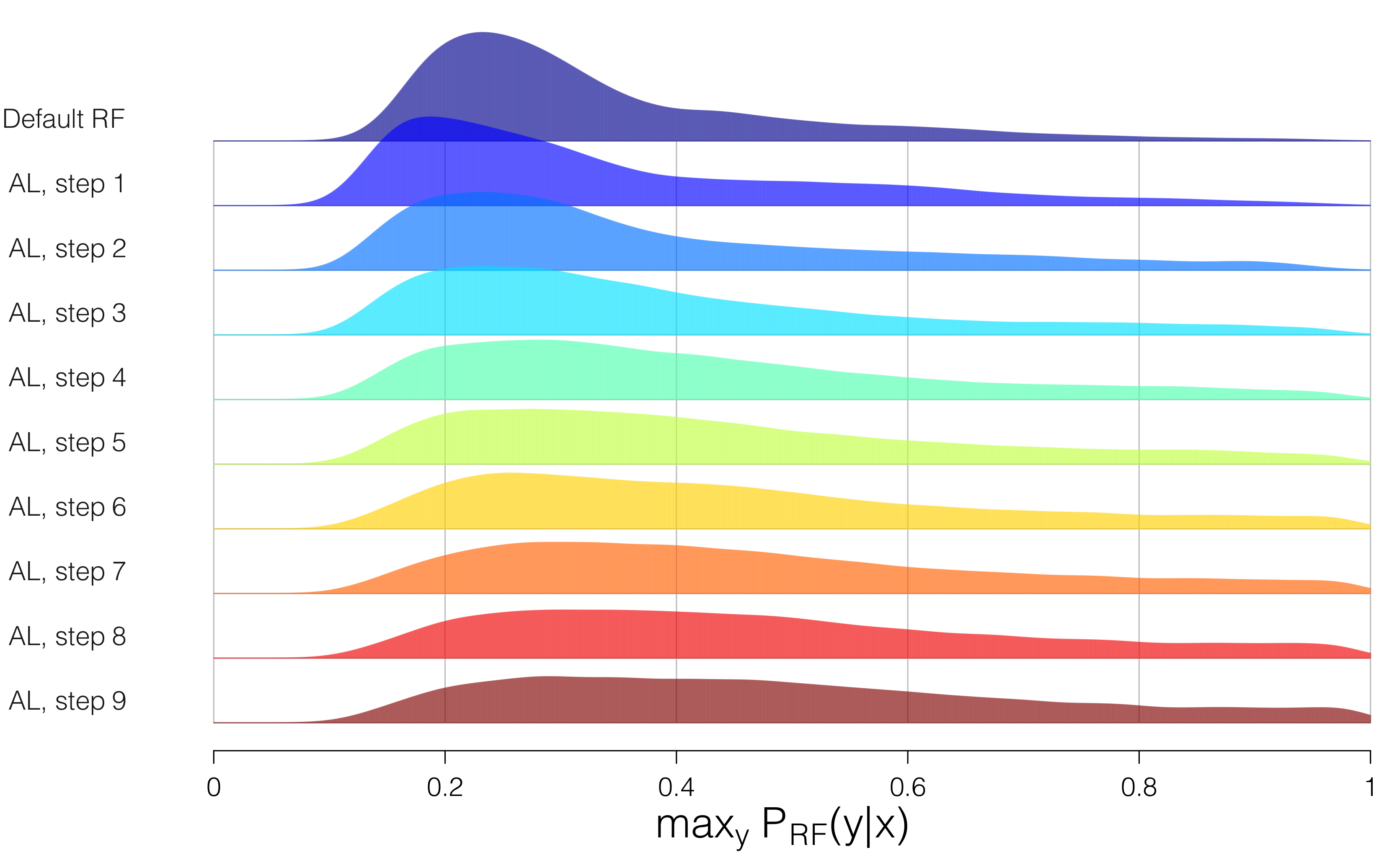}
\end{center}
\caption{ Distribution of the Random Forest $\max_y \widehat{P}_{\rm RF}(y|\x)$ values for the ASAS data, as a function of AL iteration.  For the default RF classifier, most values are smaller than 0.4, meaning that the classifier is confident on very few sources.  As the AL iterations proceed, much of the mass of the distribution gradually shifts toward larger values.  The distribution slowly becomes multimodal: for a slim majority of sources, the algorithm has high confidence, while for a substantial subset of the data the algorithm remains unsure of the classification.  \label{fig:asasprobdist}}
\end{figure}

As described in \S\ref{ss:acvs}, the default RF only attains a 65.5\% agreement with the ACVS catalog.  After 9 AL iterations, this jumps to 79.5\%, an increase of  21\% in agreement rate.  The proportion of ACVS sources in which we are confident (defined as $\max_y \widehat{P}_{\rm RF}(y|\x) > 0.5$) climbs from 14.6\% to 42.9\%.  This occurs because the selected ASAS data that are subsequently used as training data fill in sparse regions of training set feature space, thus increasing the chance that ASAS sources are in close proximity to training data and  increasing the RF maximum probabilities.  As a function of the AL iteration, both the ACVS agreement rate and the proportion of confident classifications achieved by our classifier are plotted in Figure \ref{fig:alrates}.  The full evolution of the distribution of $\max_y \widehat{P}_{\rm RF}(y|\x)$ is plotted in Figure \ref{fig:asasprobdist}.  As the iterations proceed, power is shifted  from low to high probabilities.

\begin{figure}
\begin{center}
\includegraphics[angle=0,width=6.5in]{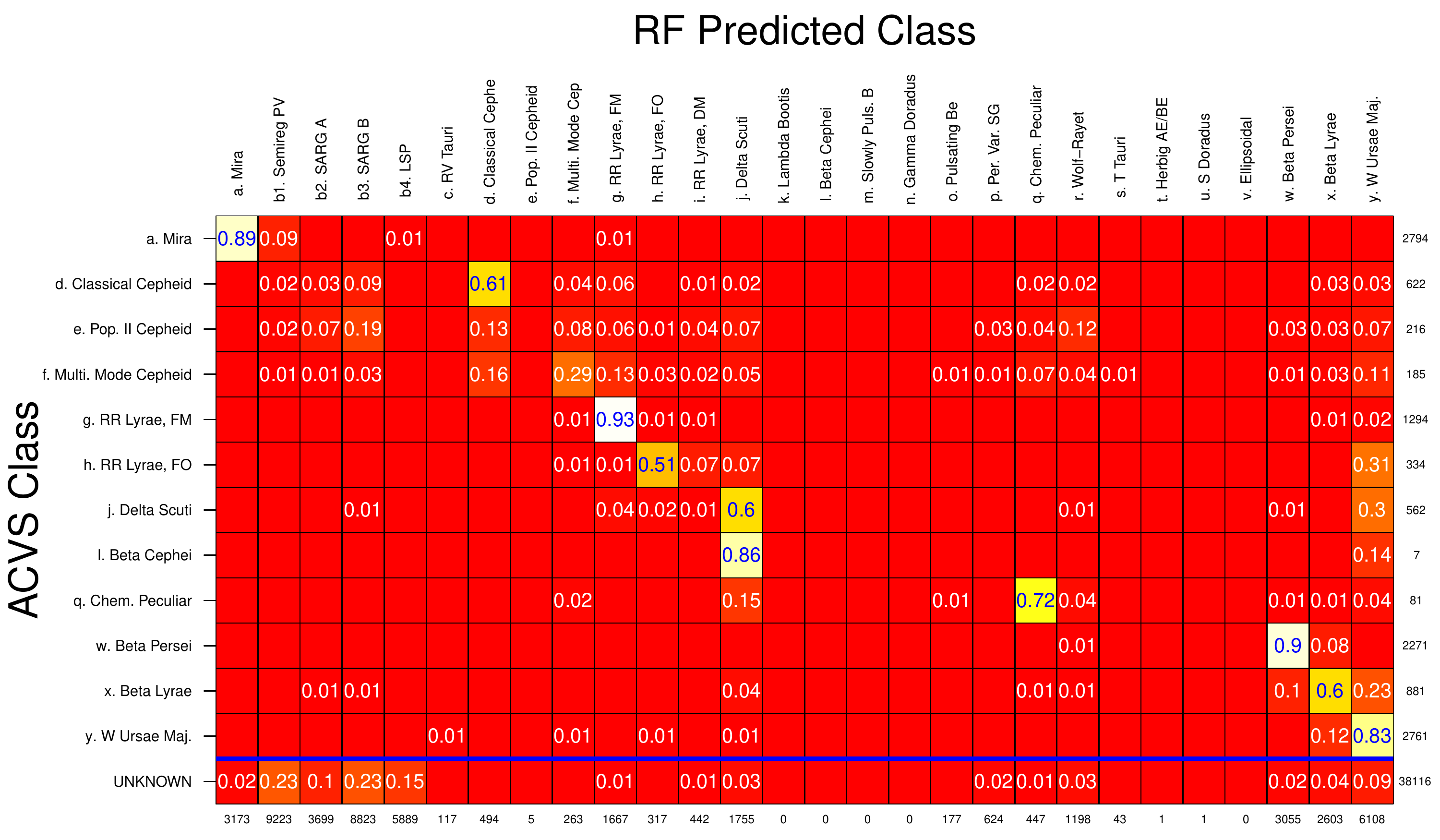}\\
\includegraphics[angle=0,width=6.5in]{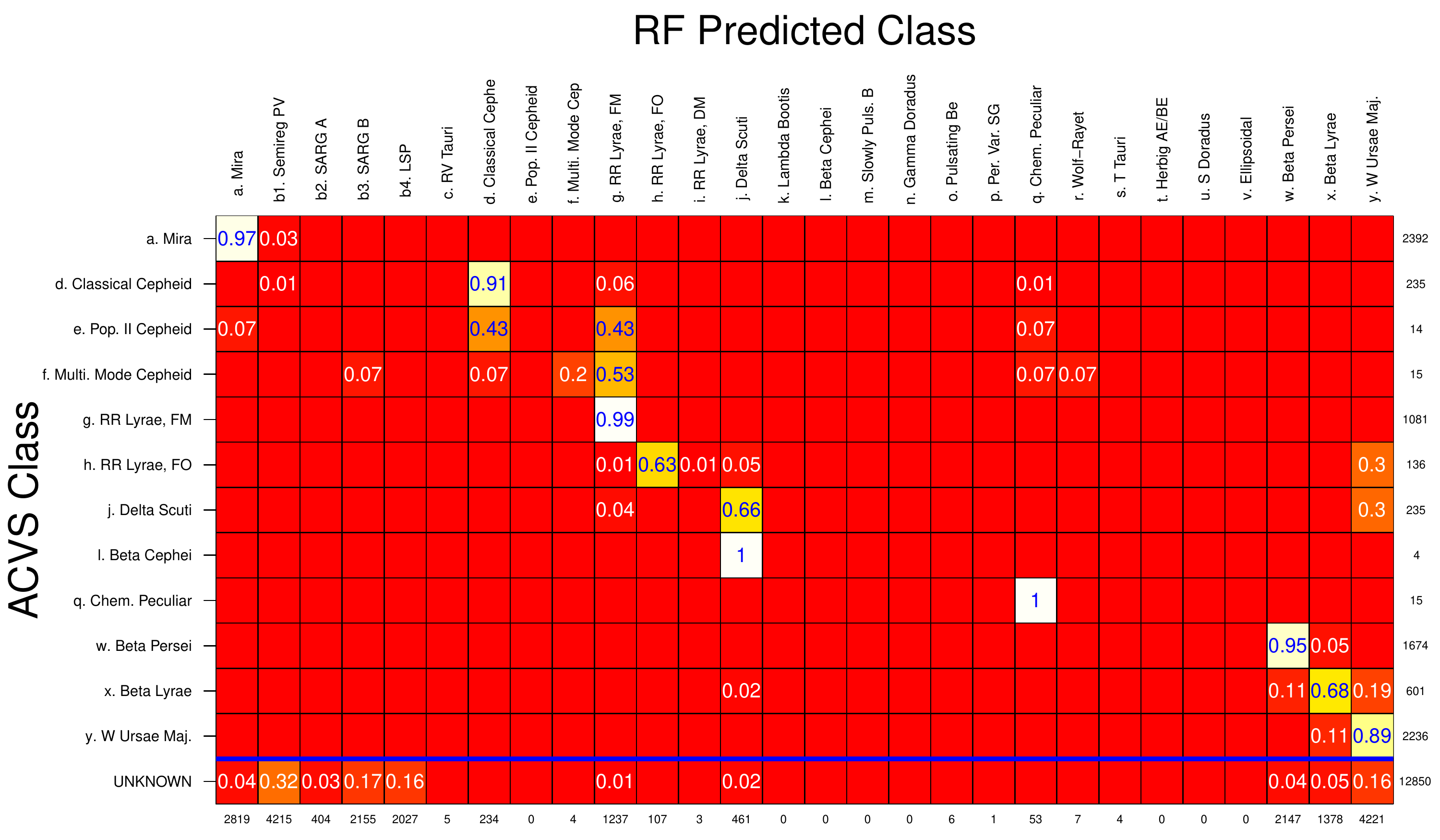}
\end{center}
\caption{Top: Classifications of the active learning RF classifier after 9 iterations of AL.  Compared to Figure \ref{fig:rfasaspred}, there is a  closer correspondence to the ACVS class labels (y axis).  Notably, the RRL, DM artifact has largely disappeared.  Bottom: Same for only sources with classification probability $> 0.5$.  Here, the agreement is even higher.  The main confusion is in classifying ACVS RR Lyrae, FO and Delta Scuti as W Ursae Maj.   \label{fig:alasaspred}}
\end{figure}

In Figure \ref{fig:alasaspred} we plot a table of the correspondence between our classifications after 9 AL iterations and the ACVS class.  Comparing to Figure \ref{fig:rfasaspred}, we see that the AL predictions more closely match the ACVS labels across most science classes.  For example, correspondence in the Classical Cepheid class raised from 24\% to 61\%, RR Lyrae, FM from 79\% to 93\%, Delta Scuti from 22\% to 60\%, and Chemically Peculiar from 1\% to 72\%.  We have also identified a number of candidates for more rare classes, such as 117 RV Tauri, 177 Pulsating Be stars, and 43 T Tauri.  Additionally, the number of RR Lyrae, DM candidates, which was artificially high for the original RF classifier, has diminished from 9109 to 442.   A summary of our ASAS classification active learning, by class, is given in Table \ref{tab:predasas}.

As a consequence of performing active learning on the ASAS data set, we were able to detect the presence of 3 additional science classes of red giant stars.  These classes were discovered by one of the AL users upon realizing that many of the queried pulsating red giant stars were low-amplitude with 10-75 day periods.  A literature search revealed that these stars naturally break into small-amplitude red giant A and B subclasses (SARG A and B, see \citealt{2004MNRAS.349.1059W}).  Furthermore, the presence of a red giant subclass of long secondary period (LSP, \citealt{2007ApJ...660.1486S}) stars was discovered and added.  Via active learning, our classifier identified 3699 SARG A, 8823 SARG B, and 5889 LSP candidates.

Our final experiment is to compare our classification results using active learning with the classification of a Random Forest that is trained on the ACVS labels.  The aim of this study is to determine whether our classifier's disagreement with ACVS is due principally to inadequacies in our classifier or mistakes and inconsistencies in the ACVS classifications.  Using a 5-fold cross-validation on the ACVS labels, a RF classifier finds a 90\% agreement rate with ACVS (compared to 79.5\% using AL).  A substantial proportion of our disagreement with ACVS results from the use of a finer taxonomy (where, e.g., we can correctly identify some of ACVS Mira candidates as Semi-Regular PVs).  Within the classes in which the AL classifier has its poorest agreement with ACVS, the ACVS RF also does not do well: for Pop. II Cepheids, the ACVS RF finds only 37\% agreement (compared to 0\%), for Multi. Mode Cepheids it finds 45\% agreement (29\%), and in Beta Cepheid it finds 0\% agreement (0\%).  This evidence points to the conclusion that the disagreement of our AL classifier to ACVS within these classes is due more to lack of self-consistency of those classes in ACVS (due either to mistakes in ACVS or absence of crucial features) than to any shortcomings in the active learning methodology.


%
%



\begin{deluxetable}{lrrrr}
\tabletypesize{\footnotesize}
\tablewidth{4.0in}
\tablecolumns{5}
\tablecaption{Results, by class, of performing active learning to classify  ASAS variable stars. \label{tab:predasas}}
\tablehead{   
 \colhead{Science Class} &
 \colhead{$N_{\rm Train}$} &
 \colhead{$N_{\rm AL Add}$\tablenotemark{a}} &
 \colhead{$N_{\rm RF}$\tablenotemark{b}} &
 \colhead{$N_{\rm AL}$\tablenotemark{c}} 
 }
\startdata
a. Mira & 144 & 20 & 3587 & 3173 \\ 
b1. Semireg PV & 42 & 59 & 5799 & 9223 \\ 
b2. SARG A & 0 & 15 & 0 & 3699 \\ 
b3. SARG B & 0 & 29 & 0 & 8823 \\ 
b4. LSP & 0 & 54 & 0 & 5889 \\ 
c. RV Tauri & 6 & 5 & 0 & 117 \\ 
d. Classical Cepheid & 191 & 16 & 324 & 494 \\ 
e. Pop. II Cepheid & 23 & 0 & 98 & 5 \\ 
f. Multi. Mode Cepheid & 94 & 4 & 162 & 263 \\ 
g. RR Lyrae, FM & 124 & 26 & 1714 & 1667 \\ 
h. RR Lyrae, FO & 25 & 14 & 51 & 317 \\ 
i. RR Lyrae, DM & 57 & 3 & 9109 & 442 \\ 
j. Delta Scuti & 114 & 19 & 822 & 1755 \\ 
k. Lambda Bootis & 13 & 0 & 0 & 0 \\ 
l. Beta Cephei & 39 & 0 & 0 & 0 \\ 
m. Slowly Puls. B & 29 & 0 & 0 & 0 \\ 
n. Gamma Doradus & 28 & 0 & 0 & 0 \\ 
o. Pulsating Be & 45 & 4 & 10 & 177 \\ 
p. Per. Var. SG & 55 & 1 & 1663 & 624 \\ 
q. Chem. Peculiar & 51 & 14 & 27 & 447 \\ 
r. Wolf-Rayet & 40 & 0 & 6683 & 1198 \\ 
s. T Tauri & 14 & 4 & 753 & 43 \\ 
t. Herbig AE/BE & 15 & 0 & 4 & 1 \\ 
u. S Doradus & 7 & 0 & 0 & 1 \\ 
v. Ellipsoidal & 13 & 0 & 0 & 0 \\ 
w. Beta Persei & 169 & 25 & 2110 & 3055 \\ 
x. Beta Lyrae & 145 & 37 & 11962 & 2603 \\ 
y. W Ursae Maj. & 59 & 66 & 5246 & 6108 \\ 
\enddata
\tablenotetext{a}{ASAS sources added to the training set after 8 AL iterations.}
\tablenotetext{b}{Number of ASAS sources classified by the default Random Forest.}
\tablenotetext{c}{Number of ASAS sources classified by the RF after 8 AL iterations.}
\end{deluxetable}

\section{Conclusions}
\label{sec:conclusions}

We have described the problem of sample selection bias (a.k.a. covariate shift) in supervised learning on astronomical data sets.  Though supervised learning has shown great promise in automatically analyzing large astrophysical databases,  care must be taken to account for the biases that occur due to distributional differences between the training and testing sets.  Here, we have argued that sample selection bias is a common problem in astronomy, primarily because the subset of well-studied astronomical objects  typically forms a biased sample of intrinsically brighter and nearby sources.  In this paper, we showed the detrimental influence of sample selection bias on the problem of supervised classification of variable stars. 

To alleviate the effects of sample selection bias, we proposed a few different methods.  We find, on a toy problem using \hipp and OGLE light curves, that active learning performs significantly better than other methods such as importance weighting, co-training, and self-training.  Furthermore, we argue that AL is a suitable method for many astronomical problems, where follow-up resources are usually available (albeit with limited availability).  Active learning simply gives a principled way to determine which sources, if followed up on, would help the supervised algorithm the most.  We show that in classifying variable stars from the ASAS survey, AL produces hugely significant improvements in performance within only a handful of iterations.  Our {\tt ALLSTARS} web interface was critical in this work, as was the participation of knowledgeable (``trained expert'') users and sophisticated crowdsourcing methods.

Though we have introduced a couple of AL querying functions, many different options are available.  In particular, we argue that the $S_2$ criterion is appropriate for our classification problem because it targets objects whose inclusion in the training set would induce the largest overall change in the classification predictions over the testing set.  However, for each problem, a different AL function will be appropriate.  The pertinent querying function depends on the problem at hand, the type of response being modeled, and the kind of supervised algorithm employed, and typically several different choices are available.

One common cause of sample selection bias in variable star classification is that data from older surveys---whose sources have typically been observed over many epochs---are commonly used to classify data from ongoing surveys, whose sources contain many fewer epochs of observation.  In addition to AL, other viable approaches to this particular problem are those of \emph{noisification}, where the training set light curves are artificially modified to resemble those of the testing set, and \emph{denoisification}, where each testing light curve is matched to a (clean) training light curve.  These techniques are currently being studied by \citet{long2011}.
 
Our discussion of sample selection bias has revolved around the use of non-parametric tools (and in particular Random Forests).  For the types of complicated classification and regression problems in astrophysics, flexible non-parametric methods are usually necessary.  However, in many applications, parametric models are appropriate.  In this parametric setting, there are several methods of overcoming sample selection bias, including Bayesian experimental design (\citealt{chaloner1995bayesian}).

We conclude by emphasizing the importance of treating sample selection bias for future petabyte-scale surveys such as Gaia and LSST.  These upcoming surveys will collect data at such massive rates that rare, unexpected, and yet-undiscovered sources will be prevalent in their data streams.  Furthermore, due to superior optics and cameras, they will probe different populations of sources than observed by any previous mission.  For these reasons, any conceivable training set constructed prior to the start of these surveys will have significant sample selection bias.  Through active learning, we now have a principled way to queue sources for targeted follow-up in order to augment  training sets to optimize the performance of machine-learned algorithms and to maximize the science that these missions produce.

\acknowledgements

 The authors acknowledge the generous support of a CDI grant (\#0941742) from the National Science Foundation.  This work was performed in the CDI-sponsored Center for Time Domain Informatics (\url{http://cftd.info}).  N.R.B. is supported through the Einstein Fellowship Program (NASA Cooperative Agreement: NNG06DO90A).  We acknowledge the help of Christopher Klein, Adam Morgan, and Brad Cenko in helping to manually classify variable stars with {\tt ALLSTARS}.  We also thank Laurent Eyer for helpful conversations.

\appendix

\section{Derivation of Active Learning Random Forest Metric}
\label{app:alrf}

In this Appendix, we derive Equation \ref{eq:alrf2} as an AL selection criterion function.  Our starting point is to select instances that maximize the total amount of change in the RF predicted probabilities of the testing data $\x \in \mathcal{U}$.  Assuming we have a labeled training set $\mathcal{L}$, the total amount of change in the testing RF probabilities due to the addition of $\x'$ to $\mathcal{L}$ is
\begin{equation}
\label{eq:app1}
S_2(\x') = \sum_{\x \in \mathcal{U}} || \PrfLx - \PrfL ||_1
\end{equation}
where we use the notation $\PrfL$ to denote the Random Forest probability that the label for instance $\x$ is $y$, where the RF is trained on the set $\mathcal{L}$.  To simplify notation, we rewrite $S_2(\x') = \sum_{\x \in \mathcal{U}} \Delta(\x',\x)$, where
\begin{eqnarray}
\label{eq:app2}
\Delta(\x',\x) &=& || \PrfLx - \PrfL ||_1\\
\label{eq:app3}
 &=& \sum_{y=1}^C | \PrfLx - \PrfL |
\end{eqnarray}
where $C$ is the total number of classes.  Equation \ref{eq:app3} follows from the definition of $\ell_1$ norm.

From Equation \ref{eq:rfprob}, $\PrfL = \frac{1}{B}\sum_b\theta_{b,\mathcal{L}}(y|\x)$, where $\theta_{b,\mathcal{L}}$ is the $b$th decision tree in the Random Forest built on training set $\mathcal{L}$.  Now, assuming that the addition of $\x'$ to $\mathcal{L}$ does not change the structure of any of the $B$ decision trees\footnote{In reality, the structure of the trees may change, but analyzing the effect on the RF of adding $\x'$ is intractable if the trees are allowed to change substantially.}, we can compute the change in the decision tree estimate in terminal node $T_b(\x')$ of tree $b$.  Let $Y(\x')$ denote the true label of source $\x'$.  In adding $\x'$ to $\mathcal{L}$, decision tree $b$ changes to
\begin{equation}
\label{eq:app4}
\TreeLx = \left\{ \begin{array}{ll}
\frac{n_b(\x') \TreeL + I(Y(\x')=y)}{n_b(\x') + 1} & \textrm{if } \x \in T_b(\x')\\
\TreeL & \textrm{if } \x \notin T_b(\x')
\end{array} \right.
\end{equation}
where $n_b(\x')$ is the number points in $\mathcal{L}$ that fall in $T_b(\x')$ and $I(\cdot)$ is a boolean indicator function.  The way to understand Equation \ref{eq:app4} is that the empirical probability estimates in the terminal node $T_b(\x')$ update to include $Y(\x')$, while the rest of the terminal nodes remain unchanged.

Therefore, if $\x \in T_b(\x')$, then the amount of change in the probability estimate is
\begin{eqnarray}
\TreeLx - \TreeL &=& \frac{n_b(\x')\TreeL + I(Y(\x')=y)}{n_b(\x')+1} - \TreeL\\
\label{eq:app6}
& = & \frac{I(Y(\x')=y) - \TreeL}{n_b(\x')+1}
\end{eqnarray}
while in all other terminal nodes of $b$, the change is 0.  

Using the result in Equation \ref{eq:app6} for tree $b$, we can compute the total amount of change, $\Delta(\x',\x)$, across the entire RF by averaging the response over the $B$ trees:
\begin{eqnarray}
\label{eq:app7}
\Delta(\x',\x) &=& \sum_{y=1}^C \left| \frac{1}{B} \sum_{b: \x \in T_b(\x')}  \frac{I(Y(\x')=y) - \TreeL}{n_b(\x')+1} \right| 
\end{eqnarray}
where $n_b(\x')$ and $\TreeL$ are quantities computed for each of the $B$ trees.  However, these entities are costly to store  for large $B$ and are not available in most  RF implementations.  To compute Equation \ref{eq:app7} directly from the standard RF output (e.g., proximity matrices and predicted probabilities), we need two approximations: (1) $n_b(\x') = \sum_{\z \in \mathcal{L}} \rho(\x',\z)$, i.e., replace the number of objects in $T_b(\x')$ by the average number over the $B$ trees, and (2) $\TreeL = \PrfL$, i.e., approximate the  probability vector of each tree by the  RF probability.  Using these approximations we have that
\begin{eqnarray}
\label{eq:app8}
\Delta(\x',\x) &\approx& \sum_{y=1}^C \left| \frac{1}{B} \sum_{b: \x \in T_b(\x')}  \frac{I(Y(\x')=y) - \PrfL}{\sum_{\z \in \mathcal{L}} \rho(\x',\z)+1} \right| \\
\label{eq:app9}
 &=&  \frac{1}{\sum_{\z \in \mathcal{L}} \rho(\x',\z)+1} \sum_{y=1}^C \left| \frac{1}{B}\sum_{b=1}^B I(\x \in T_b(\x')) \left( I(Y(\x')=y) - \PrfL\right) \right| \\
\label{eq:app10}
 &=&   \frac{1}{\sum_{\z \in \mathcal{L}} \rho(\x',\z)+1} \sum_{y=1}^C \left| I(Y(\x')=y) - \PrfL \right|  \frac{1}{B}\sum_{b=1}^B I(\x \in T_b(\x'))
\end{eqnarray}

However, we cannot directly compute this equation because do not know \emph{a priori} what the value of $Y(\x')$ is.  Luckily, we can find a lower bound on the term in Equation \ref{eq:app10} that includes $Y(\x')$, and use this to produce a \emph{conservative} estimate of $\Delta(\x',\x)$.  Our lower bound is
\begin{eqnarray*}
\sum_{y=1}^C | I(Y(\x')=y) - \PrfL | &=&  (1-\widehat{P}_{\rm{RF},\mathcal{L}}(Y(\x')|\x)) + \sum_{y \ne Y(\x')} \PrfL \\
&\ge& 1 - \widehat{P}_{\rm{RF},\mathcal{L}}(Y(\x')|\x)\\
&\ge& 1 - \max_y \PrfL
\end{eqnarray*}
Therefore, the smallest possible change in the RF probabilities is given by 
\begin{equation}
\label{eq:app11}
\Delta(\x',\x) = \frac{1 - \max_y \PrfL}{\sum_{\z \in \mathcal{L}} \rho(\x',\z)+1}  \frac{1}{B}\sum_{b=1}^B I(\x \in T_b(\x'))
\end{equation}
which is a metric that can be computed.

Now substituting the result of Equation \ref{eq:app11} into Equation \ref{eq:app1}, we have that
\begin{eqnarray}
S_2(\x') &=& \sum_{\x \in \mathcal{U}} \frac{1-\max_y \PrfL}{\sum_{\z \in \mathcal{L}} \rho(\x',\z)+1} \frac{1}{B} \sum_{b=1}^B I(\x \in T_b(\x'))\\
\label{eq:appresult}
&=& \sum_{\x \in \mathcal{U}} \frac{1-\max_y \PrfL}{\sum_{\z \in \mathcal{L}} \rho(\x',\z)+1}\rho(\x',\x)
\end{eqnarray}
which is the AL criterion, $S_2$, presented in Equation \ref{eq:alrf2}.

\bibliography{TSclassify}

\begin{thebibliography}{66}
\expandafter\ifx\csname natexlab\endcsname\relax\def\natexlab#1{#1}\fi

\bibitem[{{Auvergne} {et~al.}(2009){Auvergne}, {Bodin}, {Boisnard}, {Buey},
  {Chaintreuil}, {Epstein}, {Jouret}, {Lam-Trong}, {Levacher}, {Magnan},
  {Perez}, {Plasson}, {Plesseria}, {Peter}, {Steller}, {Tiph{\`e}ne}, {Baglin},
  {Agogu{\'e}}, {Appourchaux}, {Barbet}, {Beaufort}, {Bellenger}, {Berlin},
  {Bernardi}, {Blouin}, {Boumier}, {Bonneau}, {Briet}, {Butler}, {Cautain},
  {Chiavassa}, {Costes}, {Cuvilho}, {Cunha-Parro}, {de Oliveira Fialho},
  {Decaudin}, {Defise}, {Djalal}, {Docclo}, {Drummond}, {Dupuis}, {Exil},
  {Faur{\'e}}, {Gaboriaud}, {Gamet}, {Gavalda}, {Grolleau}, {Gueguen},
  {Guivarc'h}, {Guterman}, {Hasiba}, {Huntzinger}, {Hustaix}, {Imbert},
  {Jeanville}, {Johlander}, {Jorda}, {Journoud}, {Karioty}, {Kerjean},
  {Lafond}, {Lapeyrere}, {Landiech}, {Larqu{\'e}}, {Laudet}, {Le Merrer},
  {Leporati}, {Leruyet}, {Levieuge}, {Llebaria}, {Martin}, {Mazy}, {Mesnager},
  {Michel}, {Moalic}, {Monjoin}, {Naudet}, {Neukirchner}, {Nguyen-Kim},
  {Ollivier}, {Orcesi}, {Ottacher}, {Oulali}, {Parisot}, {Perruchot},
  {Piacentino}, {Pinheiro da Silva}, {Platzer}, {Pontet}, {Pradines},
  {Quentin}, {Rohbeck}, {Rolland}, {Rollenhagen}, {Romagnan}, {Russ}, {Samadi},
  {Schmidt}, {Schwartz}, {Sebbag}, {Smit}, {Sunter}, {Tello}, {Toulouse},
  {Ulmer}, {Vandermarcq}, {Vergnault}, {Wallner}, {Waultier}, \&
  {Zanatta}}]{2009A&A...506..411A}
{Auvergne}, M., {et~al.} 2009, \aap, 506, 411

\bibitem[{{Ball} {et~al.}(2004){Ball}, {Loveday}, {Fukugita}, {Nakamura},
  {Okamura}, {Brinkmann}, \& {Brunner}}]{2004MNRAS.348.1038B}
{Ball}, N.~M., {Loveday}, J., {Fukugita}, M., {Nakamura}, O., {Okamura}, S.,
  {Brinkmann}, J., \& {Brunner}, R.~J. 2004, \mnras, 348, 1038

\bibitem[{{Bloom} \& {Richards}(2011)}]{br11}
{Bloom}, J.~S., \& {Richards}, J.~W. 2011, {arXiv/1104.3142}

\bibitem[{Blum \& Mitchell(1998)}]{blum1998}
Blum, A., \& Mitchell, T. 1998, in Proceedings of the eleventh annual
  conference on Computational learning theory, ACM, 92--100

\bibitem[{{Bonfield} {et~al.}(2010){Bonfield}, {Sun}, {Davey}, {Jarvis},
  {Abdalla}, {Banerji}, \& {Adams}}]{2010MNRAS.405..987B}
{Bonfield}, D.~G., {Sun}, Y., {Davey}, N., {Jarvis}, M.~J., {Abdalla}, F.~B.,
  {Banerji}, M., \& {Adams}, R.~G. 2010, \mnras, 405, 987

\bibitem[{Breiman(2001)}]{2001brei}
Breiman, L. 2001, Machine learning, 45, 5

\bibitem[{Brinker(2003)}]{brinker2003}
Brinker, K. 2003, in In Proceedings of the 20th International Conference on
  Machine Learning (AAAI Press), 59--66

\bibitem[{{Butler} \& {Bloom}(2011)}]{2011butl}
{Butler}, N.~R., \& {Bloom}, J.~S. 2011, \aj, 141, 93

\bibitem[{{Carliles} {et~al.}(2010){Carliles}, {Budav{\'a}ri}, {Heinis},
  {Priebe}, \& {Szalay}}]{2010ApJ...712..511C}
{Carliles}, S., {Budav{\'a}ri}, T., {Heinis}, S., {Priebe}, C., \& {Szalay},
  A.~S. 2010, \apj, 712, 511

\bibitem[{Chaloner \& Verdinelli(1995)}]{chaloner1995bayesian}
Chaloner, K., \& Verdinelli, I. 1995, Statistical Science, 10, 273

\bibitem[{Cohn(1996)}]{cohn1996}
Cohn, D. 1996, Neural Networks, 9, 1071

\bibitem[{{Collister} \& {Lahav}(2004)}]{2004PASP..116..345C}
{Collister}, A.~A., \& {Lahav}, O. 2004, \pasp, 116, 345

\bibitem[{{D'Abrusco} {et~al.}(2007){D'Abrusco}, {Staiano}, {Longo}, {Brescia},
  {Paolillo}, {De Filippis}, \& {Tagliaferri}}]{2007ApJ...663..752D}
{D'Abrusco}, R., {Staiano}, A., {Longo}, G., {Brescia}, M., {Paolillo}, M., {De
  Filippis}, E., \& {Tagliaferri}, R. 2007, \apj, 663, 752

\bibitem[{{Debosscher} {et~al.}(2007){Debosscher}, {Sarro}, {Aerts}, {Cuypers},
  {Vandenbussche}, {Garrido}, \& {Solano}}]{2007debo}
{Debosscher}, J., {Sarro}, L.~M., {Aerts}, C., {Cuypers}, J., {Vandenbussche},
  B., {Garrido}, R., \& {Solano}, E. 2007, \aap, 475, 1159

\bibitem[{{Debosscher} {et~al.}(2009){Debosscher}, {Sarro}, {L{\'o}pez},
  {Deleuil}, {Aerts}, {Auvergne}, {Baglin}, {Baudin}, {Chadid}, {Charpinet},
  {Cuypers}, {De Ridder}, {Garrido}, {Hubert}, {Janot-Pacheco}, {Jorda},
  {Kaiser}, {Kallinger}, {Kollath}, {Maceroni}, {Mathias}, {Michel}, {Moutou},
  {Neiner}, {Ollivier}, {Samadi}, {Solano}, {Surace}, {Vandenbussche}, \&
  {Weiss}}]{2009A&A...506..519D}
{Debosscher}, J., {et~al.} 2009, \aap, 506, 519

\bibitem[{Donmez {et~al.}(2009)Donmez, Carbonell, \& Schneider}]{donmez2009}
Donmez, P., Carbonell, J., \& Schneider, J. 2009, in Proceedings of the 15th
  ACM SIGKDD international conference on Knowledge discovery and data mining,
  ACM, 259--268

\bibitem[{{Dubath} {et~al.}(2011){Dubath}, {Rimoldini}, {S{\"u}veges},
  {Blomme}, {L{\'o}pez}, {Sarro}, {De Ridder}, {Cuypers}, {Guy}, {Lecoeur},
  {Nienartowicz}, {Jan}, {Beck}, {Mowlavi}, {De Cat}, {Lebzelter}, \&
  {Eyer}}]{2011arXiv1101.2406D}
{Dubath}, P., {et~al.} 2011, {arXiv/1101.2406}

\bibitem[{{Gao} {et~al.}(2008){Gao}, {Zhang}, \& {Zhao}}]{2008MNRAS.386.1417G}
{Gao}, D., {Zhang}, Y.-X., \& {Zhao}, Y.-H. 2008, \mnras, 386, 1417

\bibitem[{Goldman \& Zhou(2000)}]{goldman2000}
Goldman, S., \& Zhou, Y. 2000, in Proceedings of the 17th International
  Conference on Machine Learning, Citeseer

\bibitem[{Heckman(1979)}]{heckman1979}
Heckman, J. 1979, Econometrica: Journal of the econometric society, 153

\bibitem[{Huang {et~al.}(2007)Huang, Smola, Gretton, Borgwardt, \&
  Scholkopf}]{huan2007}
Huang, J., Smola, A., Gretton, A., Borgwardt, K., \& Scholkopf, B. 2007,
  Advances in neural information processing systems, 19, 601

\bibitem[{{Huertas-Company} {et~al.}(2008){Huertas-Company}, {Rouan}, {Tasca},
  {Soucail}, \& {Le F{\`e}vre}}]{2008A&A...478..971H}
{Huertas-Company}, M., {Rouan}, D., {Tasca}, L., {Soucail}, G., \& {Le
  F{\`e}vre}, O. 2008, \aap, 478, 971

\bibitem[{{Kessler} {et~al.}(2010){Kessler}, {Bassett}, {Belov}, {Bhatnagar},
  {Campbell}, {Conley}, {Frieman}, {Glazov}, {Gonz{\'a}lez-Gait{\'a}n},
  {Hlozek}, {Jha}, {Kuhlmann}, {Kunz}, {Lampeitl}, {Mahabal}, {Newling},
  {Nichol}, {Parkinson}, {Philip}, {Poznanski}, {Richards}, {Rodney}, {Sako},
  {Schneider}, {Smith}, {Stritzinger}, \& {Varughese}}]{2010kess}
{Kessler}, R., {et~al.} 2010, \pasp, 122, 1415

\bibitem[{Lewis \& Gale(1994)}]{lewis1994}
Lewis, D., \& Gale, W. 1994, in Proceedings of the 17th annual international
  ACM SIGIR conference on Research and development in information retrieval,
  Springer-Verlag New York, Inc., 3--12

\bibitem[{{Lintott} {et~al.}(2008){Lintott}, {Schawinski}, {Slosar}, {Land},
  {Bamford}, {Thomas}, {Raddick}, {Nichol}, {Szalay}, {Andreescu}, {Murray}, \&
  {Vandenberg}}]{2008MNRAS.389.1179L}
{Lintott}, C.~J., {et~al.} 2008, \mnras, 389, 1179

\bibitem[{Liu(2004)}]{liu2004}
Liu, Y. 2004, Journal of chemical information and computer sciences, 44, 1936

\bibitem[{Long {et~al.}(2011)Long, El~Karoui, Rice, Richards, \&
  {Bloom}}]{long2011}
Long, J., El~Karoui, N., Rice, J., Richards, J.~W., \& {Bloom}, J.~S. 2011, in
  Preparation

\bibitem[{{LSST Science Collaborations} {et~al.}(2009){LSST Science
  Collaborations}, {Abell}, {Allison}, {Anderson}, {Andrew}, {Angel}, {Armus},
  {Arnett}, {Asztalos}, {Axelrod}, \& et~al.}]{2009arXiv0912.0201L}
{LSST Science Collaborations} {et~al.} 2009, arXiv/0912.0201

\bibitem[{{Matthews} \& {Newman}(2010)}]{2010ApJ...721..456M}
{Matthews}, D.~J., \& {Newman}, J.~A. 2010, \apj, 721, 456

\bibitem[{{Newling} {et~al.}(2011){Newling}, {Varughese}, {Bassett},
  {Campbell}, {Hlozek}, {Kunz}, {Lampeitl}, {Martin}, {Nichol}, {Parkinson}, \&
  {Smith}}]{2011MNRAS.tmp..545N}
{Newling}, J., {et~al.} 2011, \mnras, 545

\bibitem[{Nigam \& Ghani(2000)}]{nigam2000}
Nigam, K., \& Ghani, R. 2000, in Proceedings of the ninth international
  conference on Information and knowledge management, ACM, 86--93

\bibitem[{Olsson \& Tomanek(2009)}]{olsson2009}
Olsson, F., \& Tomanek, K. 2009, in Proceedings of the Thirteenth Conference on
  Computational Natural Language Learning, Association for Computational
  Linguistics, 138--146

\bibitem[{Perryman {et~al.}(1997)Perryman, Lindegren, Kovalevsky, Hoeg,
  Bastian, Bernacca, Cr{\'e}z{\'e}, Donati, Grenon, Van~Leeuwen,
  {et~al.}}]{1997perr}
Perryman, M., {et~al.} 1997, Astronomy and Astrophysics, 323, L49

\bibitem[{{Perryman} {et~al.}(2001){Perryman}, {de Boer}, {Gilmore}, {H{\o}g},
  {Lattanzi}, {Lindegren}, {Luri}, {Mignard}, {Pace}, \& {de
  Zeeuw}}]{2001A&A...369..339P}
{Perryman}, M.~A.~C., {et~al.} 2001, \aap, 369, 339

\bibitem[{{Pojmanski}(1997)}]{1997AcA....47..467P}
{Pojmanski}, G. 1997, \actaa, 47, 467

\bibitem[{{Pojmanski}(2000)}]{2000AcA....50..177P}
---. 2000, \actaa, 50, 177

\bibitem[{{Pojma{\'n}ski}(2001)}]{2001ASPC..246...53P}
{Pojma{\'n}ski}, G. 2001, in Astronomical Society of the Pacific Conference
  Series, Vol. 246, IAU Colloq. 183: Small Telescope Astronomy on Global
  Scales, ed. {B.~Paczynski, W.-P.~Chen, \&amp; C.~Lemme}, 53

\bibitem[{{Pojmanski}(2002)}]{2002AcA....52..397P}
---. 2002, \actaa, 52, 397

\bibitem[{Pojmanski {et~al.}(2005)Pojmanski, Pilecki, \& Szczygiel}]{acvs}
Pojmanski, G., Pilecki, B., \& Szczygiel, D. 2005, Acta Astronomica, 55, 275

\bibitem[{{Quadri} \& {Williams}(2010)}]{2010ApJ...725..794Q}
{Quadri}, R.~F., \& {Williams}, R.~J. 2010, \apj, 725, 794

\bibitem[{{Richards} {et~al.}(2009){Richards}, {Deo}, {Lacy}, {Myers},
  {Nichol}, {Zakamska}, {Brunner}, {Brandt}, {Gray}, {Parejko}, {Ptak},
  {Schneider}, {Storrie-Lombardi}, \& {Szalay}}]{2009AJ....137.3884R}
{Richards}, G.~T., {et~al.} 2009, \aj, 137, 3884

\bibitem[{{Richards} {et~al.}(2011{\natexlab{a}}){Richards}, {Homrighausen},
  {Freeman}, {Schafer}, \& {Poznanski}}]{2011arXiv1103.6034R}
{Richards}, J.~W., {Homrighausen}, D., {Freeman}, P.~E., {Schafer}, C.~M., \&
  {Poznanski}, D. 2011{\natexlab{a}}, arXiv/1103.6034

\bibitem[{{Richards} {et~al.}(2011{\natexlab{b}}){Richards}, {Starr}, {Butler},
  {Bloom}, {Brewer}, {Crellin-Quick}, {Higgins}, {Kennedy}, \&
  {Rischard}}]{2011rich}
{Richards}, J.~W., {et~al.} 2011{\natexlab{b}}, \apj, 733, 10

\bibitem[{Roy \& McCallum(2001)}]{roy2001toward}
Roy, N., \& McCallum, A. 2001, in Machine Learning-International Workshop,
  Citeseer, 441--448

\bibitem[{{Schulz}(2010)}]{2010ApJ...724.1305S}
{Schulz}, A.~E. 2010, \apj, 724, 1305

\bibitem[{Settles(2010)}]{sett2009}
Settles, B. 2010, Active Learning Literature Survey, Tech. rep., CS Tech. Rep.
  1648, University of Wisconsin-Madison

\bibitem[{Shimodaira(2000)}]{shimo2000}
Shimodaira, H. 2000, Journal of Statistical Planning and Inference, 90, 227

\bibitem[{{Smith} {et~al.}(2010){Smith}, {Bailer-Jones}, {Klement}, \&
  {Xue}}]{2010A&A...522A..88S}
{Smith}, K.~W., {Bailer-Jones}, C.~A.~L., {Klement}, R.~J., \& {Xue}, X.~X.
  2010, \aap, 522, A88+

\bibitem[{{Soszy{\'n}ski}(2007)}]{2007ApJ...660.1486S}
{Soszy{\'n}ski}, I. 2007, \apj, 660, 1486

\bibitem[{{Soszy{\'n}ski} {et~al.}(2011){Soszy{\'n}ski}, {Dziembowski},
  {Udalski}, {Poleski}, {Szyma{\'n}ski}, {Kubiak}, {Pietrzy{\'n}ski},
  {Wyrzykowski}, {Ulaczyk}, {Koz{\l}owski}, \&
  {Pietrukowicz}}]{2011AcA....61....1S}
{Soszy{\'n}ski}, I., {et~al.} 2011, \actaa, 61, 1

\bibitem[{Sugiyama {et~al.}(2007)Sugiyama, Krauledat, \& M{\"u}ller}]{sugi2007}
Sugiyama, M., Krauledat, M., \& M{\"u}ller, K. 2007, The Journal of Machine
  Learning Research, 8, 985

\bibitem[{Sugiyama \& M{\"u}ller(2005)}]{sugi2005}
Sugiyama, M., \& M{\"u}ller, K. 2005, Statistics \& Decisions, 23, 249

\bibitem[{{Sypniewski} \& {Gerdes}(2011)}]{2011AAS...21715004S}
{Sypniewski}, A.~J., \& {Gerdes}, D.~W. 2011, in Bulletin of the American
  Astronomical Society, Vol.~43, American Astronomical Society Meeting, 150.04

\bibitem[{Tong \& Chang(2001)}]{tong2001}
Tong, S., \& Chang, E. 2001, in Proceedings of the ninth ACM international
  conference on Multimedia, ACM, 107--118

\bibitem[{Tong \& Koller(2002)}]{tong2002}
Tong, S., \& Koller, D. 2002, The Journal of Machine Learning Research, 2, 45

\bibitem[{{Tsalmantza} {et~al.}(2007){Tsalmantza}, {Kontizas}, {Bailer-Jones},
  {Rocca-Volmerange}, {Korakitis}, {Kontizas}, {Livanou}, {Dapergolas},
  {Bellas-Velidis}, {Vallenari}, \& {Fioc}}]{2007A&A...470..761T}
{Tsalmantza}, P., {et~al.} 2007, \aap, 470, 761

\bibitem[{Tur {et~al.}(2005)Tur, Hakkani-Tur, \& Schapire}]{tur2005}
Tur, G., Hakkani-Tur, D., \& Schapire, R. 2005, Speech Communication, 45, 171

\bibitem[{{Udalski} {et~al.}(1999{\natexlab{a}}){Udalski}, {Soszynski},
  {Szymanski}, {Kubiak}, {Pietrzynski}, {Wozniak}, \& {Zebrun}}]{1999udal}
{Udalski}, A., {Soszynski}, I., {Szymanski}, M., {Kubiak}, M., {Pietrzynski},
  G., {Wozniak}, P., \& {Zebrun}, K. 1999{\natexlab{a}}, \actaa, 49, 1

\bibitem[{{Udalski} {et~al.}(1999{\natexlab{b}}){Udalski}, {Soszynski},
  {Szymanski}, {Kubiak}, {Pietrzynski}, {Wozniak}, \&
  {Zebrun}}]{1999AcA....49..223U}
---. 1999{\natexlab{b}}, \actaa, 49, 223

\bibitem[{{Udalski} {et~al.}(1999{\natexlab{c}}){Udalski}, {Soszynski},
  {Szymanski}, {Kubiak}, {Pietrzynski}, {Wozniak}, \&
  {Zebrun}}]{1999AcA....49..437U}
---. 1999{\natexlab{c}}, \actaa, 49, 437

\bibitem[{Vlachos(2008)}]{vlachos2008}
Vlachos, A. 2008, Computer Speech \& Language, 22, 295

\bibitem[{{Wadadekar}(2005)}]{2005PASP..117...79W}
{Wadadekar}, Y. 2005, \pasp, 117, 79

\bibitem[{{Wozniak} {et~al.}(2002){Wozniak}, {Udalski}, {Szymanski}, {Kubiak},
  {Pietrzynski}, {Soszynski}, \& {Zebrun}}]{2002AcA....52..129W}
{Wozniak}, P.~R., {Udalski}, A., {Szymanski}, M., {Kubiak}, M., {Pietrzynski},
  G., {Soszynski}, I., \& {Zebrun}, K. 2002, \actaa, 52, 129

\bibitem[{{Wray} {et~al.}(2004){Wray}, {Eyer}, \&
  {Paczy{\'n}ski}}]{2004MNRAS.349.1059W}
{Wray}, J.~J., {Eyer}, L., \& {Paczy{\'n}ski}, B. 2004, \mnras, 349, 1059

\bibitem[{Yan {et~al.}(2003)Yan, Yang, \& Hauptmann}]{yan2003}
Yan, R., Yang, J., \& Hauptmann, A. 2003, in Ninth IEEE International
  Conference on Computer Vision (Press), 516--523

\bibitem[{Zadrozny(2004)}]{zadr2004}
Zadrozny, B. 2004, in Proceedings of the twenty-first international conference
  on Machine learning, ACM, 114

\end{thebibliography}

\end{document}